\pdfoutput=1

\documentclass[11pt,twoside,a4paper,cmspaper,final,collab]{cms-tdr}
\def\svnVersion{477465P}\def\cmsCernNoTag{CERN-EP-2017-301}\def\cmsCernDate{\today}\def\cmsMessage{Published in the Journal of High Energy Physics as \href{http://dx.doi.org/10.1007/JHEP09(2018)148}{\doi{10.1007/JHEP09(2018)148.}}}

\begin{document}\cmsNoteHeader{EXO-17-005}

\hyphenation{had-ron-i-za-tion}
\hyphenation{cal-or-i-me-ter}
\hyphenation{de-vices}
\RCS$Revision: 477148 $
\RCS$HeadURL: svn+ssh://svn.cern.ch/reps/tdr2/papers/EXO-17-005/trunk/EXO-17-005.tex $
\RCS$Id: EXO-17-005.tex 477148 2018-10-05 13:08:03Z knam $
\newlength\cmsFigWidth
\ifthenelse{\boolean{cms@external}}{\setlength\cmsFigWidth{0.85\columnwidth}}{\setlength\cmsFigWidth{0.4\textwidth}}
\ifthenelse{\boolean{cms@external}}{\providecommand{\cmsLeft}{top\xspace}}{\providecommand{\cmsLeft}{left\xspace}}
\ifthenelse{\boolean{cms@external}}{\providecommand{\cmsRight}{bottom\xspace}}{\providecommand{\cmsRight}{right\xspace}}
\newcommand{\Hg}{\ensuremath{\PH\PGg}\xspace}
\newcommand{\zg}{\ensuremath{\PZ\gamma}\xspace}
\newcommand{\MJ}{\ensuremath{m_\mathrm{J}^\text{pruned}}\xspace}
\newcommand{\MZg}{\ensuremath{m_{\Z\gamma}}\xspace}
\newcommand{\llg}{\ensuremath{\ell\ell\gamma}\xspace}
\newcommand{\Jg}{\ensuremath{\mathrm{J}\gamma}\xspace}
\newcommand{\eeg}{\ensuremath{\Pe\Pe\gamma}\xspace}
\newcommand{\mmg}{\ensuremath{\Pgm\Pgm\gamma}\xspace}
\providecommand{\NA}{\ensuremath{\text{---}}}

\cmsNoteHeader{EXO-17-005}
\title{Search for Z$\gamma$ resonances using leptonic and hadronic final states in proton-proton collisions at $\sqrt{s}= 13$\TeV}

\date{\today}

\abstract{
A search is presented for resonances decaying to a Z boson and a photon. The analysis is based on data from proton-proton collisions at a center-of-mass energy of 13\TeV, corresponding to an integrated luminosity of 35.9\fbinv, and collected with the CMS detector at the LHC in 2016. Two decay modes of the Z boson are investigated. In the leptonic channels, the Z boson candidates are reconstructed using electron or muon pairs. In the hadronic channels, they are identified using a large-radius jet, containing either light-quark or b quark decay products of the Z boson, via jet substructure and advanced b quark tagging techniques. The results from these channels are combined and interpreted in terms of upper limits on the product of the production cross section and the branching fraction to Z$\gamma$ for narrow and broad spin-0 resonances with masses between 0.35 and 4.0\TeV, providing thereby the most stringent limits on such resonances.}

\hypersetup{%
pdfauthor={CMS Collaboration},%
pdftitle={Search for Z gamma resonances using leptonic and hadronic final states in proton-proton collisions at sqrt(s)= 13 TeV},%
pdfsubject={CMS},%
pdfkeywords={CMS, physics, Zgamma resonances}}

\maketitle
\section{Introduction}

One of the key aspects of the CERN LHC physics program is the search for new resonances predicted in theories beyond the standard model (SM). Given the fairly stringent limits already set on masses of such resonances in fermionic decay channels (e.g., via dilepton or dijet searches), it is particularly interesting to explore bosonic decay channels, which can dominate if the couplings of a new resonance to fermions are suppressed. Examples of such signatures are decays into a pair of massive bosons: VV and VH, where V represents either a $\PW$ or a $\PZ$ boson, and H refers to the recently discovered Higgs boson~\cite{Higgs1,Higgs2,Higgs3}. The latest results from these searches at the LHC are described in Refs.~\cite{diboson1,diboson2,diboson3,diboson4,diboson5,diboson6,diboson7,diboson8,diboson9,diboson10,diboson11,diboson12} for the VV channels and in Refs.~\cite{diboson7,VH1,VH2,VH3,VH4,VH5,VH6} for the VH channels.

Diboson decays involving photons, \ie, $\PW\Pgg$, $\PZ\Pgg$, and $\Pgg\Pgg$ channels, are also important, as the search in the $\Pgg\Pgg$ channel demonstrated by contributing significantly to the discovery of the Higgs boson by the ATLAS and CMS Collaborations in 2012~\cite{Higgs1,Higgs2,Higgs3}. While  a resonance decaying to diphotons cannot be a vector or an axial vector, due to the Landau--Yang theorem~\cite{Landau,Yang}, having one of the two bosons massive alleviates this constraint. Thus, charged (neutral) bosons of spin 0, 1, or 2 can be sought in the  $\PW\Pgg$ ($\PZ\Pgg$) channel, leading to a broad search program. Such resonances are predicted in many extensions of the SM, such as technicolor~\cite{EL} and little Higgs~\cite{FS} models, as well as models with an extended Higgs boson sector~\cite{BT,LLS} or with extra spatial dimensions~\cite{DHR,ASS}.

In this paper, we describe a search for $\PZ\gamma$ resonances in the leptonic ($\ell\ell$, where $\ell$ refers to $\Pe$ or $\mu$) and hadronic
decay channels of the \PZ boson, as well as the combination of these channels. While the results in this paper are interpreted in terms of spin-0 resonances, they are broadly applicable to spin-1 and spin-2 states, as the signal acceptance depends only weakly on the spin of the resonance~\cite{Aaboud:2017uhw}. Similar searches in a combination of leptonic and hadronic decay channels of the \PZ boson have been recently published by ATLAS~\cite{ATLAS-Vgamma3} at $\sqrt{s} = 13$\TeV and by CMS at $\sqrt{s} = 8$ and 13\TeV~\cite{Zgamma-had}, based on significantly smaller integrated luminosities. Other searches for $\PZ\Pgg$ resonances have been performed only in the dilepton channels. These include searches by the L3 Collaboration at the CERN LEP~\cite{L3} and the D0 Collaboration at the Fermilab Tevatron~\cite{D0-1,D0-2}. At the LHC, they have been done by ATLAS at $\sqrt{s} = 7,$ 8, and 13\TeV~\cite{ATLAS-Vgamma1,ATLAS-Vgamma2,Aaboud:2017uhw}, and CMS at $\sqrt{s} = 8$ and 13\TeV~\cite{Zgamma-lep}, as well as by ATLAS and CMS using the combined 7 and 8\TeV data~\cite{CMS-HZg,ATLAS-HZg}, and by ATLAS using the 13\TeV data~\cite{Aaboud:2017uhw} in the context
of a search for the $\PH\to \PZ\gamma$ decay.

The present search is for a resonance with a relatively narrow width appearing as an excess over the smooth \zg invariant mass (\MZg) spectrum constructed from an energetic photon
and the $\PZ\to\ell\ell$ or $\PZ\to\qqbar$ decay products.
While a search in the leptonic channels
has lower SM backgrounds, resulting in higher sensitivity to resonance masses
$<$1\TeV, at larger mass values, where backgrounds are small, the hadronic channels, with their higher branching
fraction, dominate the sensitivity. The backgrounds in both channels are determined directly from fits to data.

The \PZ boson decays in the leptonic channels
are reconstructed using an electron or a muon pair.
The dominant backgrounds in the $\ell\ell\Pgg$ channel are the irreducible contribution from
continuum $\PZ\Pgg$  production and the reducible backgrounds from either
final-state radiation in $\PZ \to \ell\ell$ events or  from \PZ boson production in association with one or more jets ($\PZ$+jets),
where a jet is misidentified as a photon.

The \PZ boson decay products in the hadronic channels can be
reconstructed either as two well-separated small-radius jets, or as a single large-radius jet (J) resulting from the merging of the two quark jets because of the large Lorentz boost of a \PZ boson produced in the decay of a heavy resonance.
The fraction of events corresponding to the merged topology, which has low background from SM sources,
increases with the mass of the resonance. To optimize signal relative to background, in this paper we consider just the merged 
jet topology (\Jg), and thus the search in the hadronic channels is focused on relatively high resonance masses where the trigger for these 
events is efficient. We use jet substructure techniques to infer the presence of two subjets, and dedicated b tagging algorithms to identify those subjets that originate from b quark fragmentation. This provides the means to distinguish a signal from the dominant
background from prompt-photon and QCD multijet production, with one of the jets spuriously passing
jet substructure requirements (and, in the latter case, with another jet mimicking a photon).

\section{The CMS detector}

The central feature of the CMS apparatus is a superconducting solenoid of 6\unit{m} internal diameter, providing a magnetic field of 3.8\unit{T}. Within the solenoid volume are a silicon pixel and strip tracker, a lead tungstate crystal electromagnetic calorimeter (ECAL), and a brass and scintillator hadron calorimeter (HCAL), each composed of a barrel and two endcap sections. Forward calorimeters extend the pseudorapidity coverage provided by the barrel and endcap detectors. Muons are detected in gas-ionization chambers embedded in the steel flux-return yoke outside the solenoid.

The silicon tracker measures charged particles within the pseudorapidity range $\abs{\eta} < 2.5$. It consists of 1440 silicon pixel and 15\,148 silicon strip detector modules. For nonisolated particles of $1 < \pt < 10\GeV$ and $\abs{\eta} < 1.4$, the track resolutions are typically 1.5\% in \pt and 25--90 (45--150)\mum in the transverse (longitudinal) impact parameter \cite{TRK-11-001}

The ECAL consists of 75\,848 lead tungstate crystals, which provide coverage in pseudorapidity $\abs{\eta} < 1.48 $ in a barrel region (EB) and $1.48 < \abs{\eta} < 3.0$ in two endcap regions (EE). Preshower detectors consisting of two planes of silicon sensors interleaved with a total of $3 X_0$ of lead are located in front of each EE detector.

In the region $\abs{\eta} < 1.74$, the HCAL cells have widths of 0.087 in pseudorapidity and 0.087 in azimuth ($\phi$). In the $\eta$-$\phi$ plane, and for $\abs{\eta} < 1.48$, the HCAL cells map on to $5 \times 5$ arrays of ECAL crystals to form calorimeter towers projecting radially outwards from close to the nominal interaction point. For $\abs{\eta } > 1.74$, the coverage of the towers increases progressively to a maximum of 0.174 in $\Delta \eta$ and $\Delta \phi$. Within each tower, the energy deposits in ECAL and HCAL cells are summed to define the calorimeter tower energies, subsequently used to provide the energies and directions of hadronic jets.

The electron momentum is estimated by combining the energy measurement in the ECAL with the momentum measurement in the tracker. The momentum resolution for electrons with $\pt \approx 45\GeV$ from $\Z \rightarrow \Pe \Pe$ decays ranges from 1.7\% for nonshowering electrons in the barrel region to 4.5\% for showering electrons in the endcaps~\cite{CMS:EGM-13-001}.

Muons are measured in the pseudorapidity range $\abs{\eta} < 2.4$, with detection planes made using three technologies: drift tubes, cathode strip chambers, and resistive-plate chambers. Matching muons to tracks measured in the silicon tracker results in a relative transverse momentum resolution for muons with $20 <\pt < 100\GeV$ of 1.3--2.0\% in the barrel and better than 6\% in the endcaps. The \pt resolution in the barrel is better than 10\% for muons with \pt up to 1\TeV~\cite{CMS-PAPER-MUO-10-004}.

Events of interest are selected using a two-tiered trigger system~\cite{Khachatryan:2016bia}. The first level, composed of custom hardware processors, uses information from the calorimeters and muon detectors to select events at a rate of around 100\unit{kHz} within a time interval of less than 4\mus. The second level, known as the high-level trigger, consists of a farm of processors running a version of the full event reconstruction software optimized for fast processing, and reduces the event rate to around 1\unit{kHz} before data storage.

A more detailed description of the CMS detector, together with a definition of the coordinate system used and the relevant kinematic variables, can be found in Ref.~\cite{Chatrchyan:2008zzk}.

\section{Data sets and event selection}

The data used in this search correspond to an integrated luminosity of 35.9\fbinv recorded by the CMS experiment at $\sqrt{s} = 13$\TeV in 2016.
The high instantaneous luminosity delivered by the LHC results in additional interactions in the same or neighboring bunch crossings (pileup) as the hard scattering interaction. The average number of pileup interactions in the 2016 data set is around 23.

In the \eeg channel, the selected events are required to pass a double-photon trigger with the transverse momentum
$\pt > 60\GeV$ and pseudorapidity $\abs{\eta} < 2.5$ requirements on both photon candidates. Since the photon trigger requirements do not include any track veto, this trigger is equally efficient in selecting photon and electron candidates. A combination of single-muon triggers requiring $\pt > 50\GeV$ and
$\abs{\eta} < 2.4$ on a muon candidate are used in the \mmg channel.
In the \Jg channel, we require a logical ``OR" of several triggers with the separate requirements: the scalar sum of transverse energies of all reconstructed jets ($\HT$) is above 800 or 900\GeV; a jet is present with the transverse energy above 500\GeV; and a photon candidate is present with $\pt>165$ or 175\GeV and $\abs{\eta} < 2.5$. We determine the selection efficiency for these trigger combinations
using unbiased data samples collected with different triggers. The triggers are found to be 98--100\% efficient with respect to the offline selection, for the entire mass range used, in all three channels. The small residual inefficiency is taken into account when calculating the signal acceptance.

Simulated signal events of spin-0 resonances decaying to $\Z\gamma$ are generated at leading order (LO) in perturbative QCD
using \PYTHIA 8.205~\cite{Sjostrand:2007gs} with the CUETP8M1~\cite{Monash,CUETP8M1} underlying-event tune.
Several samples are generated with masses ranging from 0.3 to 4.0\TeV. Two resonance width assumptions were used in the simulation:
one, termed ``narrow", has its width ($\Gamma_\mathrm{X}$) set to 0.014\% of the resonance mass ($m_\mathrm{X}$), and the second, referred to as ``broad", has $\Gamma_\mathrm{X}/m_\mathrm{X} = 5.6\%$. The first choice corresponds to a resonance with
a natural width much smaller than the detector resolution. The second choice facilitates a direct comparison with the previous CMS publications~\cite{Zgamma-lep,Zgamma-had}. We assume no interference between signal and the SM nonresonant $\PZ\gamma$ production.

Simulated background events do not enter the analyses directly, as the backgrounds are obtained from fits to
data, but are used to assess the accuracy of the background model
and to optimize event selection.
Standard model nonresonant $\PZ(\ell\ell)\Pgg$ production, which is expected to be the dominant background process in the \llg channel,
is generated at next-to-leading order (NLO) accuracy using the \MGvATNLO 2.3.3 generator~\cite{MADGRAPH5, aMCNLO}.
The $\PZ(\ell\ell)$+jets events with a jet misidentified as a photon, which constitute a subdominant source of background, are generated at LO
using \MGvATNLO, as are the dominant $\Pgg$+jets and QCD multijet events,
as well as subdominant hadronically decaying $\PW$+jets and $\PZ$+jets backgrounds in the \Jg channel.
All background events are processed with \PYTHIA for the description of fragmentation and hadronization.

All simulated samples were produced using NNPDF3.0~\cite{Ball:2014uwa} parton distribution functions (PDFs), processed with the full CMS detector model based on \GEANTfour~\cite{Geant}, and reconstructed with the same suite of programs as used for collision data. Pileup effects are taken into account by superimposing minimum bias events on the hard scattering interaction. The simulated samples are reweighted to match the reconstructed vertex multiplicity distribution observed in data.

The particle-flow (PF) event algorithm~\cite{PFlow} aims to reconstruct and identify each individual particle in an event, based on an optimized combination of information from the various elements of the CMS detector. The energy of photons is directly obtained from the ECAL measurement, corrected for zero-suppression effects. The energy of electrons is determined from a combination of the electron momentum at the primary interaction vertex as determined by the tracker, the energy of the corresponding ECAL cluster, and the energy sum of all bremsstrahlung photons spatially compatible with originating from the electron track. The energy of muons is obtained from the curvature of the reconstructed muon track. The energy of charged hadrons is determined from a combination of their momentum measured in the tracker and the matching ECAL and HCAL energy deposits, corrected for zero-suppression effects and for the response function of the calorimeters to hadronic showers. Finally, the energy of neutral hadrons is obtained from the corresponding corrected ECAL and HCAL energy deposits.

The events must contain at least one reconstructed primary vertex with at least four associated tracks,
with transverse (longitudinal) coordinates required to be within 2\,(24)\unit{cm} of the nominal collision point.
The reconstructed vertex with the largest value of summed physics-object $\pt^2$ is taken to be the primary interaction vertex. The physics objects are the jets, clustered using the jet finding algorithm~\cite{Cacciari:2008gp,Cacciari:2011ma} with the tracks assigned to the vertex as inputs, and the associated missing transverse momentum, taken as the negative vector sum of the transverse momenta of those jets.

Electron candidates must pass loose identification criteria based on the shower shape variables, 
on the ratio of energy deposits in the associated HCAL and ECAL cells,
on the geometrical matching between the energy deposits and the associated track,
and on the consistency between the energy reconstructed in the calorimeter and the momentum measured in the tracker~\cite{CMS:EGM-13-001}.

Muon candidates are reconstructed from tracks found in the muon system that are associated with the tracks
in the inner tracking systems. One muon candidate is required to pass a loose identification~\cite{Sirunyan:2018fpa}.
Another muon candidate is required to pass a tighter identification based on
the numbers of associated hits found in the pixel and strip trackers, on the numbers of hits and track segments in the muon detector,
and on criteria for the matching between the silicon detector track and the muon track segments~\cite{Sirunyan:2018fpa}. 

Leptons are required to be isolated from other energy deposits in the event. This is expected for signal leptons from $\PZ$  boson decays, but is not the case for backgrounds from nonprompt leptons originating, e.g., from $\PQb$ hadron decays. The relative isolation of a lepton is defined
as the scalar sum of the transverse momenta of all relevant PF candidates within a cone around the lepton, divided by the \pt of the lepton candidate. For an electron, the cone size $\Delta R = \sqrt{(\Delta\eta)^2 + (\Delta\phi)^2}$ depends on its \pt:
\begin{linenomath}
\begin{equation}\begin{aligned}
	\Delta R &=\begin{cases}
		0.2, & \text{for } \pt \le 50\GeV,\\
		\frac{10\GeV}{\pt}, & \text{for } 50 < \pt \le 200\GeV, \text{ and} \\
		0.05, & \text{for } \pt > 200\GeV.
\end{cases}\label{eq:miniIso}
\end{aligned}
\end{equation}
\end{linenomath}
The electron isolation is based on the photons, and charged and neutral hadrons found in the isolation cone. Charged hadrons originating from pileup vertices are excluded from the sum. The contribution to the isolation sum from neutral pileup particles is accounted for by using the average energy density method~\cite{Cacciari:2007fd}. The varying isolation cone radius in Eq. (\ref{eq:miniIso}) takes into account the aperture of $\PQb$ hadron decays as a function of their \pt, and reduces the inefficiency from accidental overlap of electrons from $\PZ$ boson decays and jets. For muons, a fixed cone of a size $\Delta R = 0.3$ is used, and the isolation is based on all charged-particle tracks within the isolation cone, excluding the candidate muon track. In the case of two spatially close muons in the event, with overlapping isolation cones, both muons are excluded from each isolation sum. This procedure, together with the use of a variable cone size for electron isolation,
ensures high lepton identification efficiency even in the topologies where a Z boson has a high Lorentz boost, as expected for Z bosons produced in a decay of a heavy resonance. The relative isolation of electron and muon candidates is required to be less than 0.1.

Photon identification is based on a multivariate analysis, employing
a boosted decision tree algorithm~\cite{BDT1,BDT2}. The inputs to the algorithm include
shower shape variables, isolation sums computed from PF candidates in a cone of radius $\Delta R = 0.3$ around the photon candidate, and variables that account for the dependences of the shower shape and isolation variables on the pileup
~\cite{CMS-EGM-14-001}. In addition, a conversion-safe electron veto~\cite{CMS-EGM-14-001} is applied.
Photon candidates are required to pass a working point that corresponds to a typical photon reconstruction and identification efficiency of 90\%, in
the photon $\pt$ range used in the analysis.

In the \Jg channel, large-cone jets are used to reconstruct hadronically decaying highly Lorentz boosted $\PZ$ boson candidates.
Jets are reconstructed from PF candidates clustered using the anti-\kt
algorithm~\cite{Cacciari:2008gp} with a distance parameter of 0.8. Charged hadrons not originating from the primary vertex are not considered in the jet clustering. Corrections based on the jet area~\cite{Cacciari:2007fd} are applied to remove the energy contribution of neutral hadrons from pileup interactions. The energies of the jets are further corrected for the response function of the calorimeter.
These corrections are extracted from simulation and confirmed with in situ measurements using the energy balance in dijet, multijet, $\Pgg$+jet, and leptonically decaying $\PZ$+jet events~\cite{Chatrchyan:2011ds,Khachatryan:2016kdb}.
Additional quality criteria are applied to jets to remove rare spurious noise patterns in the calorimeters, and also to suppress leptons misidentified as jets. The jet energy resolution amounts typically to 15\% at 10\GeV, 8\% at 100\GeV, and 4\% at 1\TeV. 
Jets must have $\pt > 200$\GeV and $\abs{\eta} < 2.0$.
The requirement on the jet $\eta$ suppresses background from $\gamma$+jets and QCD multijet events, and
ensures that the core of the jet is within the tracker volume of the detector ($\abs{\eta} < 2.5$).
The latter requirement is important for subsequent b quark tagging.

Events in the \llg channel are required to have two same-flavor leptons (electrons or muons) and a photon.
Additionally, leptons in the \mmg channel are required to have opposite electric charge. This requirement is not used in the \eeg channel due to a nonnegligible probability to misreconstruct the charge of an electron candidate because of an energetic bremsstrahlung. The leading electron (muon) is required to have $\pt > 65$\,(52)\GeV and $\abs{\eta} < 2.5$\,(2.4). The subleading lepton is required to have $\pt > 10\GeV$ and to be found in the same pseudorapidity range as the leading lepton. The photon in the \eeg (\mmg) channel is required to satisfy $\pt > 65$\,(40)\GeV and $\abs{\eta} < 2.5$. Electrons and photons in the ECAL barrel-endcap transition region ($1.44 < \abs{\eta} < 1.57$) are excluded from the analysis. In the \eeg channel, the \pt thresholds on the electrons and photons in the ECAL endcap region are increased to 70\GeV, in order to ensure a fully efficient trigger. Photons are required to be separated from lepton candidates by $\Delta R > 0.4$, to reduce the background from final-state radiation in $\PZ \to \ell\ell$ events. The invariant mass of the dilepton system is required to be $50 < m_{\ell\ell} < 130\GeV$. The minimum dilepton mass requirement suppresses contributions from $\Pp\Pp\to\gamma\gamma^*$ events, where an internal conversion
of a photon produces a lepton pair. Finally, we require the ratio of the photon \pt to $\MZg$ to be greater than 0.27.
This requirement suppresses backgrounds due to jets misidentified as photons,
without significant loss in the signal efficiency and without introducing a bias
in the $\MZg$ spectrum. We search for resonances in the $\MZg$ spectrum above 300 (250)\GeV in the electron (muon) channel.

For the \Jg channel, the photon candidates are required to have $\pt > 200$\GeV and to fall within the barrel fiducial region of the ECAL ($\abs{\eta} < 1.44$). Events with a photon reconstructed in the endcap region suffer from high $\Pgg$+jets background and do not add to the sensitivity of the analysis; therefore they are not considered. Photon candidates in the event are required to be separated from large-radius jets by
a distance of $\Delta R > 1.1$, which guarantees that the photon isolation cone is not contaminated with the jet constituents.

To identify $\PZ$ boson candidates in the \Jg channel, the reconstructed large-radius jet mass, evaluated after applying a jet pruning
algorithm~\cite{Ellis:2009su,Ellis:2009me}, is used. The jet pruning reclusters the jet constituents and eliminates soft, large-angle QCD radiation, which otherwise contributes significantly to the jet mass. The pruning algorithm reclusters each jet starting from its original constituents with the Cambridge--Aachen (CA) algorithm~\cite{CA} and discards soft and wide-angle recombinations in each step of the iterative CA
procedure. The pruned jet mass ($\MJ$) is computed from the sum of the four-momenta of the remaining constituents, which are corrected with the same factor as has already been used in the generic jet reconstruction described above. A detailed description of the pruning algorithm can be found in Ref.~\cite{Khachatryan:2014vla}. For the signal selection, we require a $\PZ$ candidate to have $75 < \MJ < 105$\GeV.

Finally, in the \Jg channel a requirement is imposed on the ratio of photon $\pt$ to the reconstructed $\zg$ mass of $\pt/\MZg > 0.34$, with the cutoff chosen, based on a study of simulated signals and backgrounds, to maximize the discovery potential for a narrow resonance. We search for resonances with masses $\MZg > 650$\GeV in this channel.

To further discriminate against the QCD multijet and $\Pgg+$jets backgrounds in the \Jg channel, we categorize the events according to the likelihood of a large-radius jet to contain subjets originating from b quark fragmentation and to contain exactly two subjets. In order to do so, we employ subjet b tagging and $N$-subjettiness~\cite{Nsubjettiness} variables ($\tau_N$). The $N$-subjettiness observable measures the spatial distribution of jet constituents relative to candidate subjet axes in order to quantify how well the jet can be divided into $N$ subjets. Subjet axes are determined by a one-pass optimization procedure, which minimizes $N$-subjettiness~\cite{subjet-axes}. In particular, the ratio of 2- to 1-subjettiness, $\tau_{21} = \tau_2/\tau_1$, offers an excellent separation between the QCD jets and jets from vector boson decays~\cite{JME-16-003}, which tend to have lower $\tau_{21}$ values than the former.

To infer the presence of b quark subjets within a large-radius jet, pruned jets are split into two subjets by reversing the final iteration of the jet clustering algorithm.
These subjets are classified according to the probability of their originating from b quarks, based on results from
the combined secondary vertex (CSVv2) b tagging algorithm~\cite{BTV,Sirunyan:2017ezt}.
A jet is identified as being consistent with the $\PZ\to\bbbar$ decay
when at least one of its subjets satisfies the medium operating point of the CSVv2 algorithm, and the other subjet
satisfies the loose operating point. The medium and loose operating points
correspond to 70 and 85\% (20 and 50\%) in the b jet tagging efficiency for $\pt < 300\GeV$ ($\pt=1\TeV$),
and 1--2\% and 10--15\% misidentification probability of a light-flavor jet, respectively.
If an event contains a $\PZ\to\bbbar$ candidate, it is classified as ``b tagged". For the rest of the events, if the large-radius jet has $\tau_{21} < 0.45$, we classify the event as ``$\tau_{21}$ tagged". Otherwise, the event is assigned to the ``untagged" category. These three categories are mutually exclusive and are combined for the final result. The additional classification according to the $\tau_{21}$ value enhances signal sensitivity by 10--15\% at low to intermediate signal masses (up to $\sim$2\TeV), relative to a previous analysis in the hadronic channels~\cite{Zgamma-had}.

\section{Background and signal modeling}
\subsection{Background modeling}
\label{sec:bkg}

Simulations in the \llg channels indicate that 80--90\% of the background remaining after
the full event selection is from SM $\PZ$ boson production accompanied by initial-state photon radiation,
with the remainder mostly from $\PZ$+jets events, with a jet misreconstructed as a photon.
The background $\MZg$ distributions fall steeply and smoothly with increasing mass, in these channels.
Likewise, studies for the \Jg channel based on simulated background samples and on the
lower sideband of the jet mass distribution ($50 < \MJ < 70\GeV$) in data show that the
invariant mass distribution \MZg of the SM background also falls smoothly in this channel,
and that the distributions of kinematic observables
derived from the lower jet mass sideband match those for the signal selection.

The background is measured directly in data, through unbinned
maximum-likelihood fits to the observed $\MZg$ distributions, performed separately in each channel.
The background in each channel is parametrized with an empirical function.
Different families of functions inspired by the ones used in searches for beyond-the-SM phenomena in the dijet, multijet, diphoton, and VV channels at hadron colliders are evaluated in the signal region using simulation (\llg channels) or the lower jet mass sideband (\Jg channel). Examples of these functions are: $f(x) = P_0(1-x^{1/3})^{P_1}/x^{P_2+P_3(\ln x)}$~\cite{Aad:2015mzg}, $g(x) = P_0(1-x)^{P_1}/x^{P_2 + P_3(\ln x)}$~\cite{Aaltonen:2008dn}, $h(x) = P_0 x^{P_1}\exp(P_2 x + P_3 x^2)$~\cite{UA2-1}, where $x = \MZg/\sqrt{s}$, $\sqrt{s}$ is the center-of-mass energy (13\TeV), and the number of the fit parameters $P_i$  shown is the maximum order considered. The choice of the order within a family of fitting functions for the background distribution is made independently in each channel using the Fisher $F$-test~\cite{F-test}, which balances the quality of the fit against the number of parameters. The choice among the families of functions is optimized based on the results of the bias test described below. The same function~\cite{Khachatryan:2016hje} was chosen in both the \llg and \Jg channels:
\begin{linenomath}
\begin{equation}
\frac{\rd{}N}{\rd\MZg} = P_0\,x^{P_1+P_2(\ln x)},
\label{eq:Func}
\end{equation}
\end{linenomath}
where $P_0$ is a normalization parameter,
and $P_1$, $P_2$ describe the shape of the invariant mass distribution.

\begin{figure}[htpb]
  \centering
	    \includegraphics[width=0.45\textwidth]{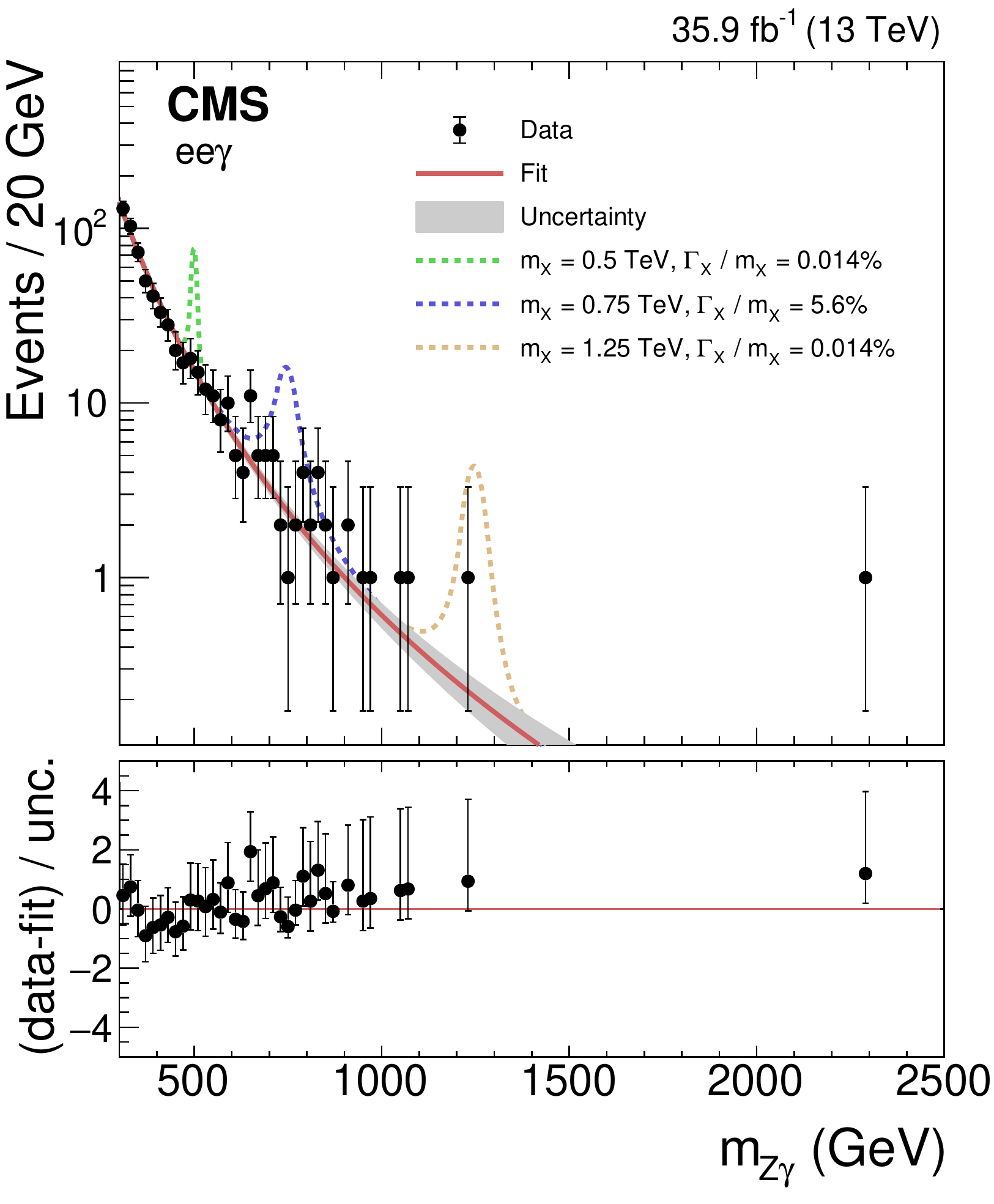}
	    \includegraphics[width=0.45\textwidth]{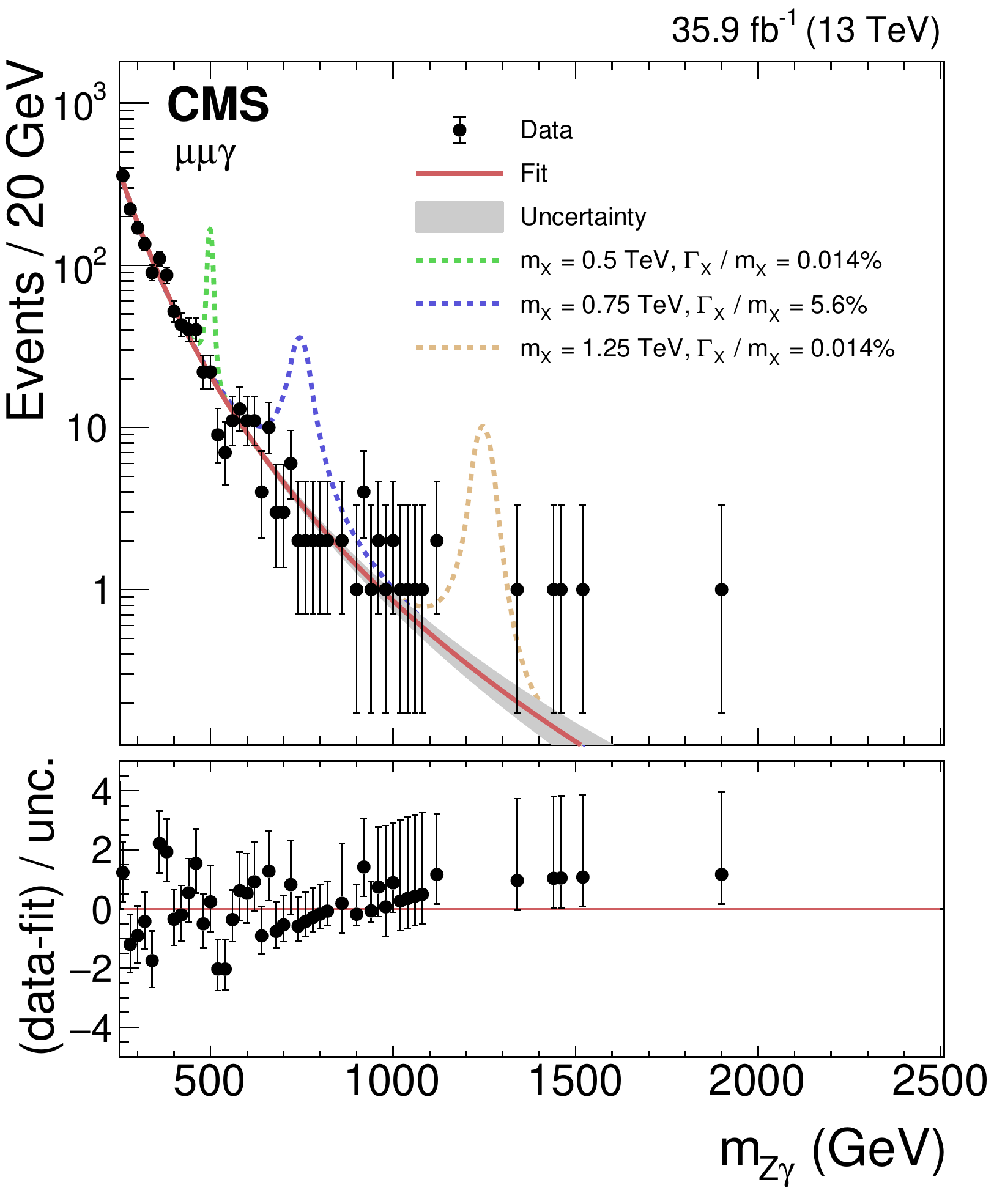}
 \caption{Observed $\MZg$ invariant mass spectra
in the $\eeg$~(left)  and $\mmg$~(right) channels.
The best fits to the background-only hypotheses are represented by the red lines, with their 68\% CL uncertainty bands given by the gray shadings. Several narrow and broad signal benchmarks with arbitrary normalization are shown on top of the background prediction with the dashed lines. The lower panels show the difference between the data and the fits,
divided by the uncertainty, which includes the statistical uncertainties in the data and the
fit. For bins with a small number of entries, the error bars correspond to the Garwood confidence intervals~\protect\cite{Garwood}.}
    \label{fig:13tev_fits}
\end{figure}

In the \llg channels, the absence of significant bias in the fit to background is verified by generating a  large number of pseudo-experiments using the simulated background shapes, fitting them with different background models, and measuring the difference between the input and fitted background yields in various $\MZg$ windows within the entire search range.
A pull variable is defined in each window by the difference between the input and fitted yields, divided by the combined statistical uncertainties in the data and the fit.
If the absolute value of the median in the pull distribution is found to be $>$0.5, an additional uncertainty is assigned to
the background parametrization. A modified pull distribution is then constructed, increasing
the statistical uncertainty by an extra term, denoted as the bias term.
The bias term is parametrized as a smooth function of $\MZg$ and tuned to make the absolute
value of the median of the modified pull distribution to be $<$0.5 in all mass windows.
This additional uncertainty is included in the likelihood function by adding to the background model
a component with a distribution that is the same as the signal, but a normalization coefficient distributed
as a Gaussian of mean zero, and a width equal to the integral of the bias term over the signal mass window, defined as the full width at half maximum.
The inclusion of this additional component takes into account the possible
mismodeling of the background shape.
The bias term in the \llg analysis corresponds to ${\approx}0.3\unit{events}/\GeVns{}$ at $\MZg = 400$\GeV and smoothly falls to
${\approx}2\times 10^{-4}$ events/GeV around $\MZg=2\TeV$.
The observed $\MZg$ invariant mass spectra in data are shown in Fig.~\ref{fig:13tev_fits} for the $\eeg$ (left) and $\mmg$ (right) channels.
The results of the fits and their uncertainties at 68\% confidence level (CL) are shown by the lines and the shaded bands, respectively.

Similarly, in the \Jg channel, the mass
spectra in the three analysis categories, derived either from the low jet mass sideband in data
or from simulated background samples, are fitted with a variety of alternative functions to generate
pseudo-data sets. Additionally,
in a set of pseudo-experiments, signals with different mass values and cross sections close to the
expected 95\% CL limits are injected. The full spectra are fitted with the chosen
function of Eq.~(\ref{eq:Func}) together with a signal model (discussed in Section~\ref{sec:signal}), and the signal
cross section is extracted. Distributions of the difference between the data and the fits, divided by the overall uncertainty for the obtained signal
cross section, are constructed, and their shapes are found to be consistent
with a normal distribution with a mean less than 0.5 and a width
consistent with unity. 

\begin{figure}[h!t]
\centering
\includegraphics[width=0.32\textwidth]{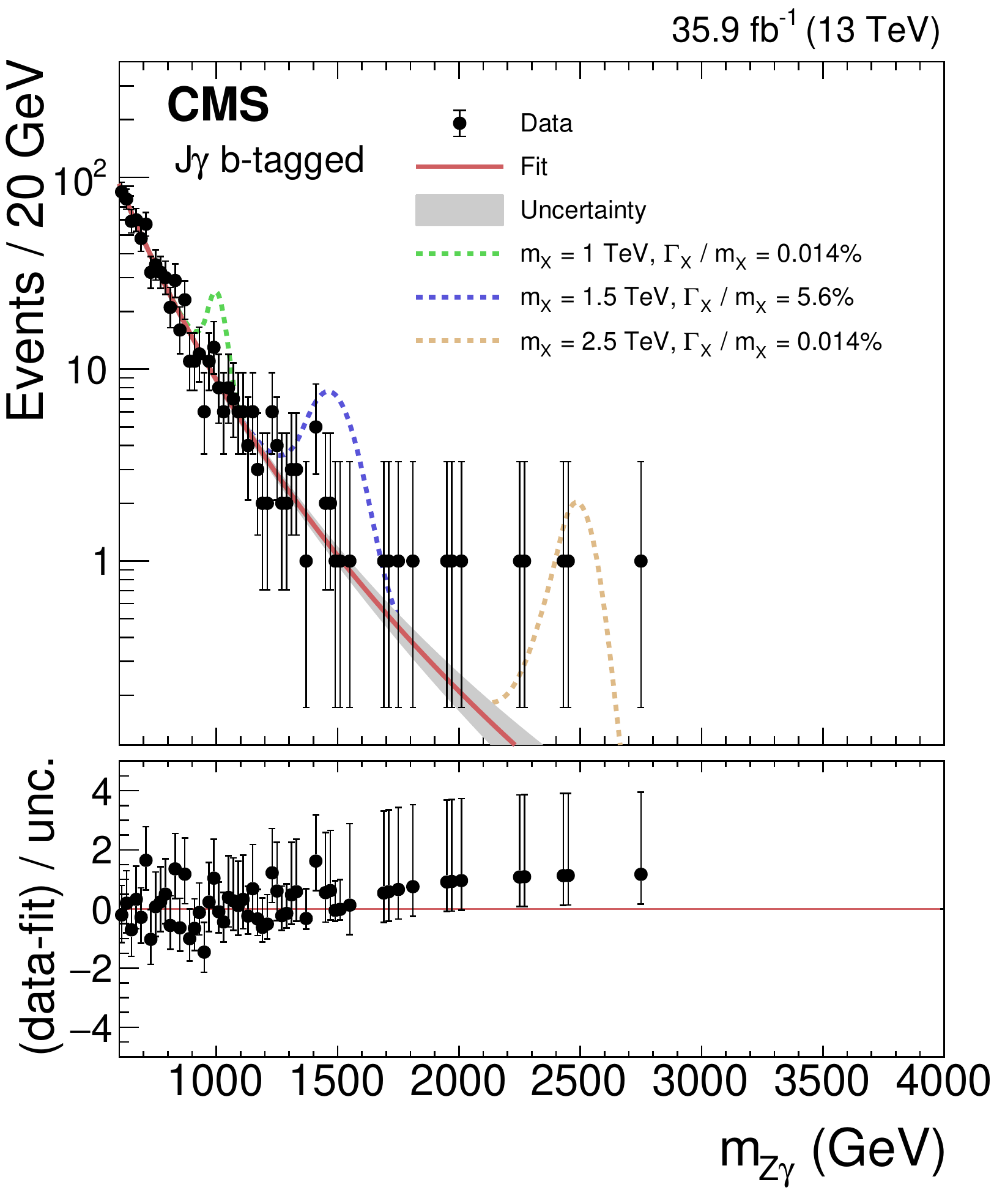}
\includegraphics[width=0.32\textwidth]{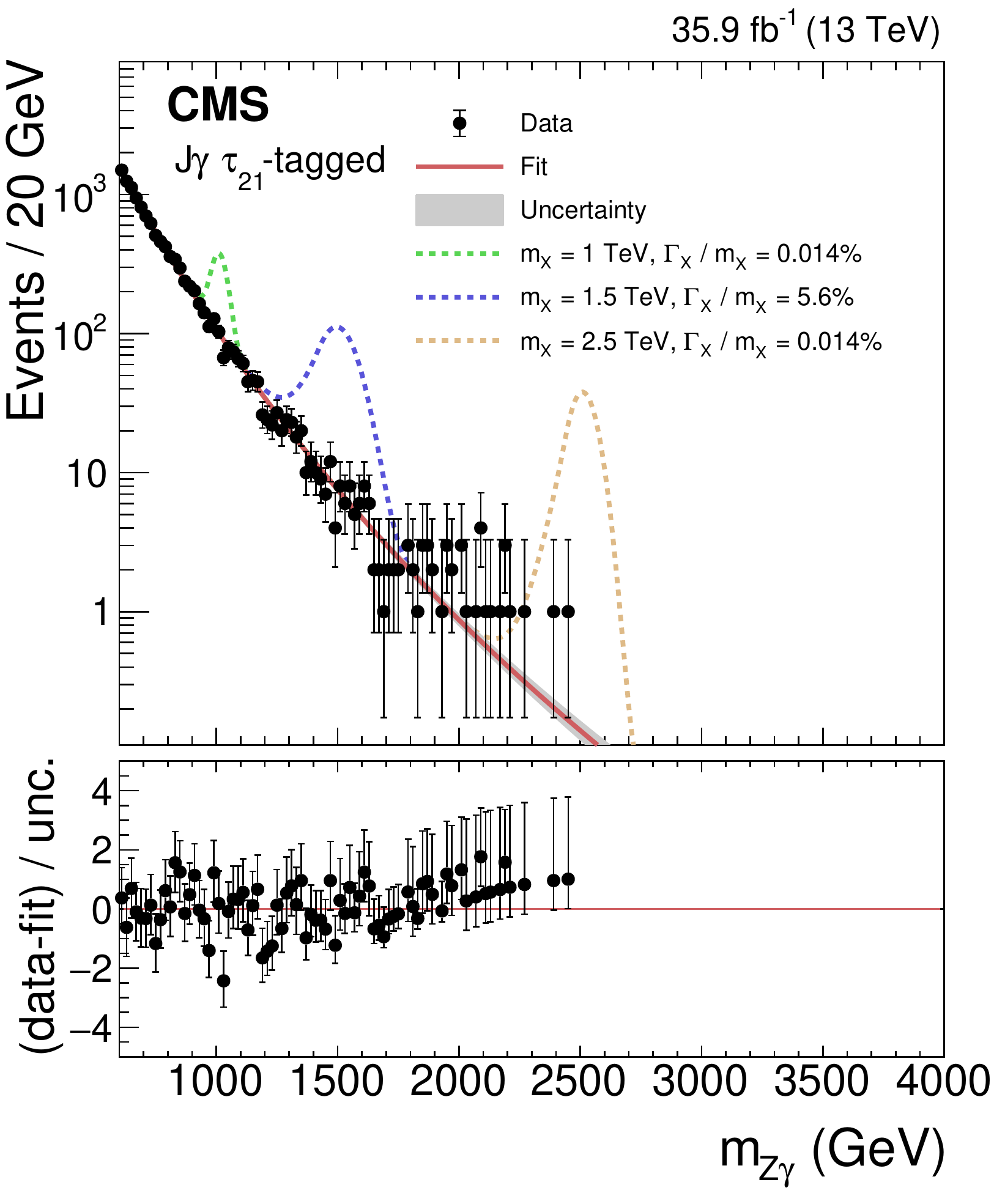}
\includegraphics[width=0.32\textwidth]{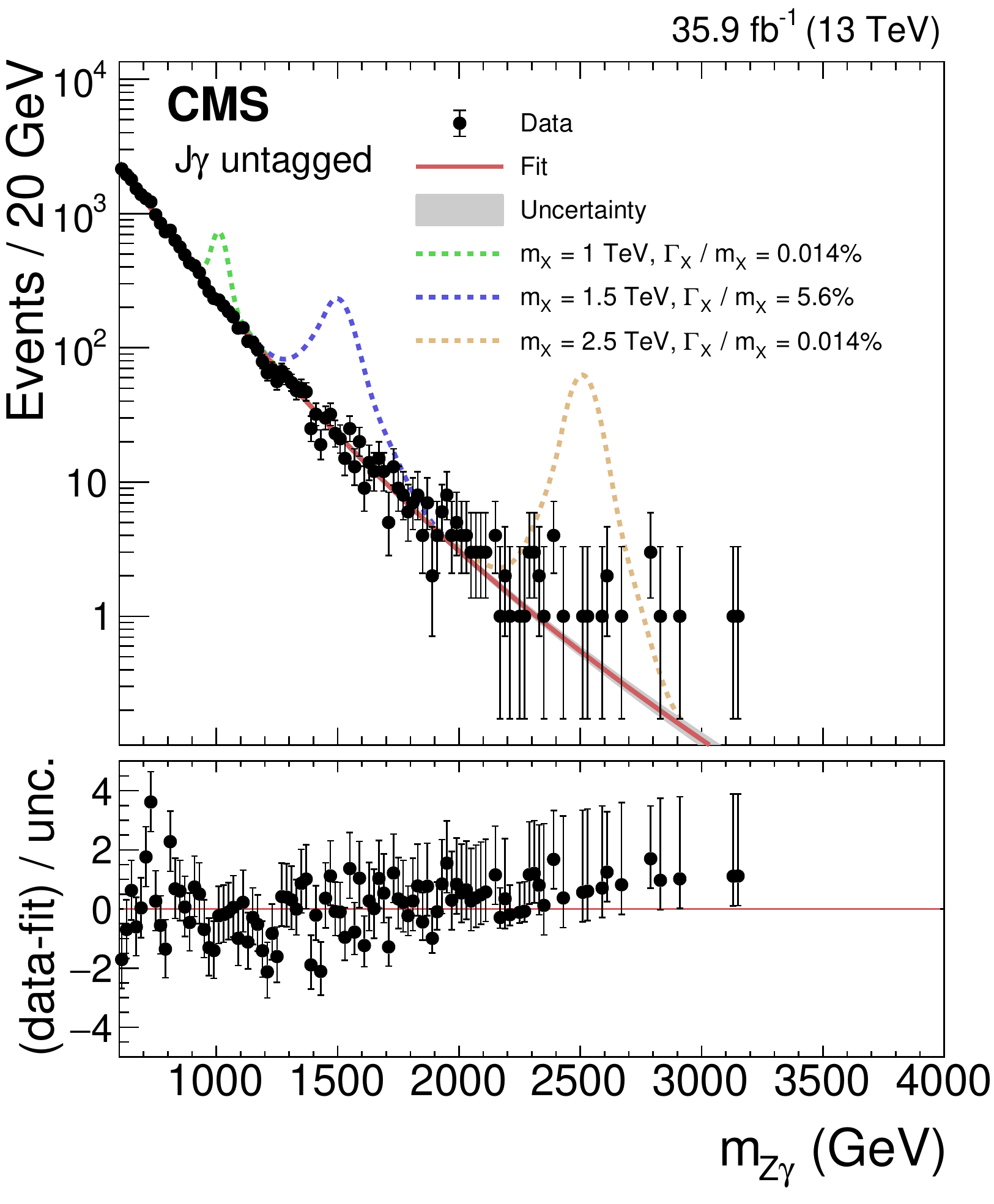}
\caption{Observed $\MZg$ invariant mass spectra
in the \Jg channel in the b-tagged (left), $\tau_{21}$-tagged (center), and untagged (right) categories.
The best fits to the background-only hypotheses are represented by the red lines, with their 68\% CL uncertainty bands given by the gray shadings. Several narrow and broad signal benchmarks with arbitrary normalization are shown on top of the background prediction with the dashed lines. The lower panels show the difference between the data and the fits,
divided by the uncertainty, which includes the statistical uncertainties in the data and the
fit. For bins with a small number of entries, the error bars correspond to the Garwood confidence intervals~\protect\cite{Garwood}.}
\label{fig:bkgfitsSR}

\end{figure}

Thus, any possible systematic bias from the choice of the functional form in the region of low background is 
proven to be small compared to the statistical uncertainty from the accuracy of the measurements. 
In the region of large background, the uncertainty in the signal efficiency (discussed in Section~\ref{sec:syst}) completely 
dominates the effect of the background uncertainty on the limits. Thus we assign the statistical uncertainty in the fits as the only 
uncertainty in the background predictions.
The observed $\MZg$ invariant mass spectra in the signal region ($75 < \MJ < 105\GeV$)
for the b-tagged, $\tau_{21}$-tagged, and untagged categories,
along with the corresponding fits, are shown in Fig.~\ref{fig:bkgfitsSR}.

\subsection{Signal modeling}
\label{sec:signal}

The signal distribution in $\MZg$ is obtained from the generated events that pass the full selection.
The signal shape is parametrized with a Gaussian core and two power-law tails,
namely an extended form of the Crystal Ball (CB) function~\cite{CrystalBallRef}. We find this functional form to provide an adequate description for both narrow and broad signals in the entire mass range used in the analysis. To derive the signal shapes for
the intermediate mass values where simulation points are not available, a linear morphing~\cite{morphing} of the shapes obtained from
the simulation is used. The typical mass resolution for narrow-width signal events is 1\% for the $\eeg$ channel,
1--2\% for the $\mmg$ channel, depending on the mass of the resonance, and 3\% in the \Jg channel.

The product of signal acceptance and efficiency for a narrow resonance in the $\eeg$ ($\mmg$) channel
rises from about 27\,(42)\% at
$\MZg = 0.35$\TeV to about 46\,(55)\% at $\MZg = 2$\TeV, and remains steady until 4\TeV. In the \Jg channel, the product of signal acceptance and efficiency for narrow resonances increases
from 7\,(3)\% at 0.65\TeV to 9\,(9)\% at 4\TeV in the untagged ($\tau_{21}$-tagged) category,
and is between 2 and 3\% for the b-tagged category for the resonance masses between 0.65 and 4\TeV.

For a broad resonance the product of signal acceptance and efficiency is similar to the narrow-resonance case up to 2\TeV. At large resonance masses ($>$2\TeV), however, the effect of rapidly falling PDFs introduces a lower tail in the signal mass distribution. The exact characteristics of this tail are quite sensitive to the resonant line shape. We therefore truncate the mass distribution of the resonance to correspond to the core of the line shape, defined as a window centered on the maximum of the CB function with a width given by $\pm$5 times the CB function parameter $\sigma$, describing the standard deviation of its Gaussian core. The tails outside this window are conservatively discarded in the signal acceptance calculations and when fitting the data. As a result, the product of signal acceptance and efficiency falls to 2\% at 4\TeV for the $\eeg$ and $\mmg$ events, to 0.2\% in the untagged and $\tau_{21}$-tagged categories, and to $<$0.1\% in the b-tagged category.

\section{Systematic uncertainties}
\label{sec:syst}

The statistical uncertainty in the fits of the background function to data is taken for the background uncertainty in all channels.

The following systematic uncertainties in the signal are defined below, and summarized in Table~\ref{tab:systSummary}:
\begin{itemize}
\item {Integrated luminosity:} the uncertainty in the CMS integrated luminosity
is based on cluster counting in the silicon pixel detector
and amounts to 2.5\%~\cite{LUM-17-001}.
\item {PDFs:} a 1.0--3.5\% uncertainty in the signal efficiency
    that takes into account the variation in the kinematic acceptance of the
    analysis is estimated using replicas of the NNPDF3.0 set, following the PDF4LHC prescription~\cite{PDF4LHC}. The uncertainty in the signal cross section due to the PDF choice is not considered.
\item{Pileup:} the uncertainty due to the pileup description in the signal simulation, evaluated by changing the total inelastic cross section governing the average multiplicity of pileup interactions by $\pm$5.0\%~\cite{inelastic}, translates to a 1.0\% uncertainty in the signal acceptance in all channels.
\item {Trigger:} the uncertainty due to the trigger efficiency differences in
data and simulation in the \llg analysis is estimated with dedicated studies with leptons from $\PZ$ boson decays and amounts to 1.0\,(3.5)\% for
the \eeg~(\mmg) channel. In the \Jg channel, a 2.0\% uncertainty covers the variation of the trigger efficiency across the mass range probed in the analysis.
\item {Photon efficiency:} the systematic uncertainty due to the
differences in the photon identification efficiency between data and simulation is evaluated
with $\PZ\to \Pe\Pe$ events in which the electrons are used as proxies for photons,
and amounts to 1.5\%~\cite{CMS-EGM-14-001}.
\item {Lepton efficiency:} the systematic uncertainty due to the
differences in the lepton identification efficiency in data and simulation is evaluated
with $\PZ\to \Pe\Pe$ ($\mu\mu$) events and amounts to 2.5\,(2.0)\%
in the \eeg (\mmg) channel.
\item {b tagging efficiency:} the uncertainty due to the difference in the b tagging efficiency in data and simulation is estimated from control samples in data and simulation enriched in b quarks~\cite{Sirunyan:2017ezt}, and translates into a 15--32\% uncertainty in the signal yield in the \Jg channel. It is anticorrelated between the b-tagged and the other two categories, as it induces signal migration between the categories.
\item {{${\tau}_{21}$} tagging efficiency:} to account for the difference between the $\tau_{21}$ distributions in data and simulation, a scale factor of $0.97 \pm 0.06$~\cite{JME-16-003} is introduced for simulated signal samples. This translates into an uncertainty of 10--12\% in the signal yield in the $\tau_{21}$-tagged category and is anticorrelated with that in the untagged category.
\item {Electron and photon energy scale and resolution:} the electron and photon energy scale is known with 0.1--5.0\%
precision, depending on the energy.  This uncertainty is based on the accuracy of the
    energy scale at the $\PZ$ boson peak and its extrapolation to higher
    masses, and translates into a 0.2--4.6 (0.1--2.3)\% correlated uncertainty in the \MZg
    scale in the \eeg channel (\mmg and \Jg channels). The uncertainty in the electron and photon energy resolution based on the Gaussian smearing evaluated at the \PZ boson peak translates to a 10 (5)\% uncertainty in the \MZg resolution in the \eeg channel (\mmg and \Jg channels).
 \item {Muon momentum scale and resolution:} the muon momentum scale is measured with 0.1--5.0\%
    precision up to $\pt = 200$\GeV, with an additional 0.1--6.0\% uncertainty at higher values, resulting in a 0.1--4.6\% uncertainty in the
    \MZg mass scale in the \mmg channel. A 10\% uncertainty in the \MZg resolution in the \mmg channel is conservatively assigned to account for the uncertainty in the muon momentum resolution.
\item {Jet energy scale (JES), jet mass scale (JMS), jet energy resolution (JER), and jet mass resolution (JMR):} the uncertainties~\cite{Chatrchyan:2011ds,Khachatryan:2016kdb,Khachatryan:2014vla} are propagated to all the relevant quantities, and affect both the signal yield and its shape. The overall effect of these uncertainties added in quadrature corresponds to approximately 5.0\% uncertainty in the signal yield, as determined by changing the four-momenta of the jets accordingly and carrying out the full analysis with the modified quantities.
\end{itemize}

\begin{table}[h!t]
\centering
\topcaption{Summary of the systematic uncertainties in the signal yield (upper part of the table) or shape (lower part of the table). A dash indicates that the uncertainty does not apply.}
\label{tab:systSummary}
\begin{tabular}{l*{5}c}
\hline
Source             & \eeg & \mmg & b-tagged & $\tau_{21}$-tagged & Untagged \\\hline
Integrated luminosity         & 2.5\% & 2.5\% & 2.5\% & 2.5\% & 2.5\% \\
PDFs                & 1--3.5\%   & 1--3.5\%   & 1--3.5\%   & 1--3.5\%   & 1--3.5\%   \\
Pileup             & 1\%   & 1\%   & 1\% & 1\% & 1\% \\
Trigger            & 1\%   & 3.5\% & 2\%   & 2\%   & 2\%   \\
Photon efficiency  & 1.5\% & 1.5\% & 1.5\% & 1.5\% & 1.5\% \\
Lepton efficiency  & 2.5\% & 2\% & \NA & \NA & \NA \\
b tagging efficiency   & \NA & \NA & 15--32\% & anticorr. & anticorr. \\
$\tau_{21}$ tagging efficiency   & \NA & \NA & \NA & 10--12\% & anticorr. \\[2ex]
e/$\Pgg$ energy scale  & 0.2--4.6\% & 0.1--2.3\% & 0.1--2.3\% & 0.1--2.3\% & 0.1--2.3\% \\
e/$\Pgg$ energy resolution & 10\% & 5\% & 5\% & 5\% & 5\% \\
Muon momentum scale   & \NA & 0.1--4.6\% & \NA & \NA & \NA \\
Muon momentum resolution   & \NA & 10\% & \NA & \NA & \NA \\
JES and JER         & \NA & \NA &  3.2\% & 3.2\% & 3.2\% \\
JMS and JMR         & \NA & \NA & 4.1\% & 4.1\% & 4.1\%\\
\hline
\end{tabular}
\end{table}
\section{Results}

The data are consistent with the background-only expectations in all channels. We set upper limits on the production cross section of heavy spin-0 resonances using the asymptotic approximation~\cite{Gross} of the modified frequentist CL$_\mathrm{s}$ method~\cite{Junk:1999kv,Read:2002hq,ATL-PHYS-PUB-2011-011}, with a likelihood ratio used as a test statistic, and uncertainties incorporated as nuisance parameters with log-normal (normalization) or Gaussian (shape) priors. The limits are set in the mass range between 0.35\,(0.30) and 4.0\TeV in the \eeg (\mmg) channel and 0.65--4.0\TeV in the \Jg channel. We note that the asymptotic approximation tends to produce lower cross section limits than the exact CL$_\mathrm{s}$ calculations in the regions with low background. We tested that the difference is at most 4 (30\%) for resonance masses below 1 TeV (around 3 TeV). 

\subsection{The \texorpdfstring{\llg}{ll gamma} channels}

\begin{figure}[htb!]
  \centering
      \includegraphics[width=0.48\textwidth]{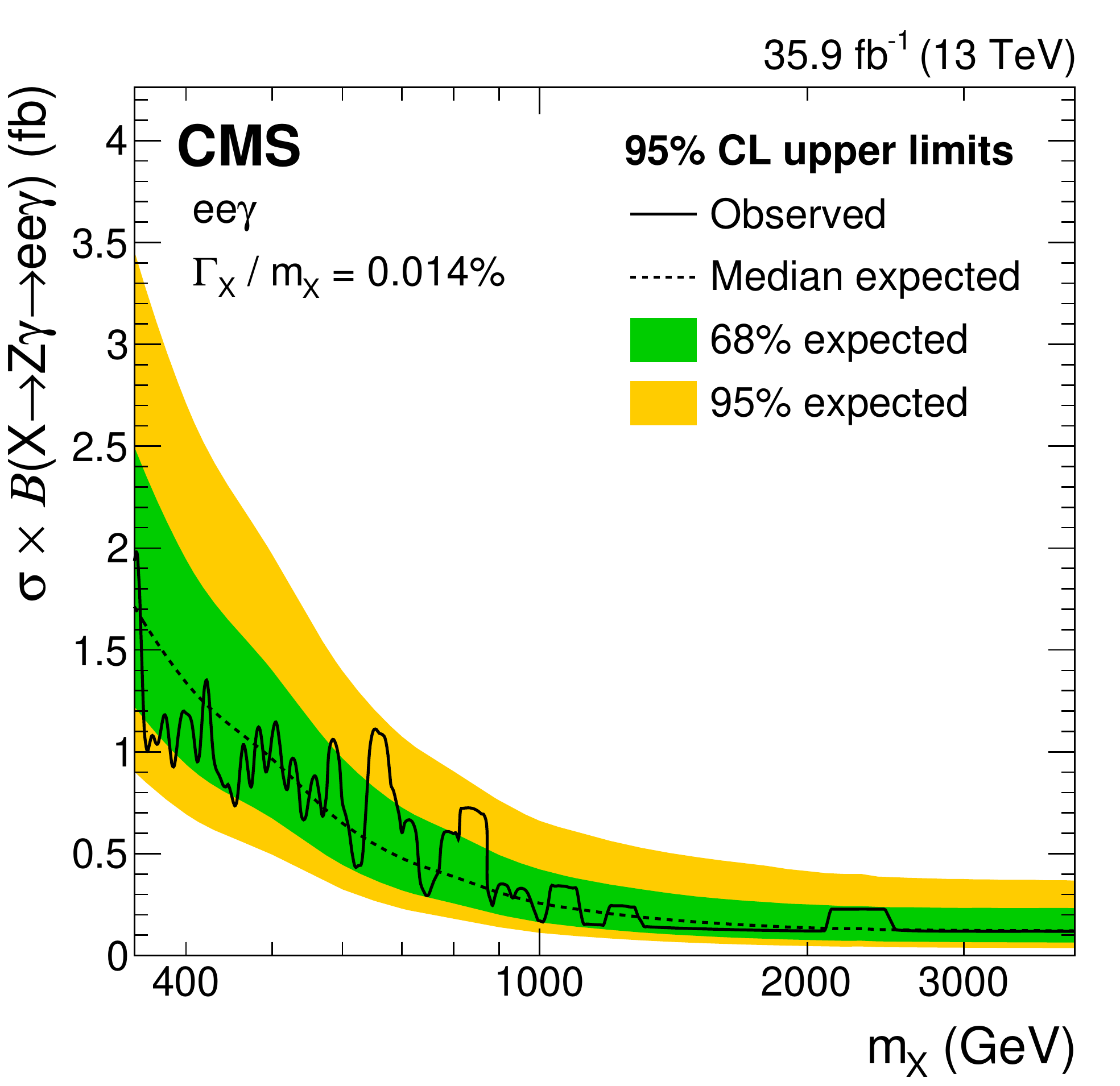}
      \includegraphics[width=0.48\textwidth]{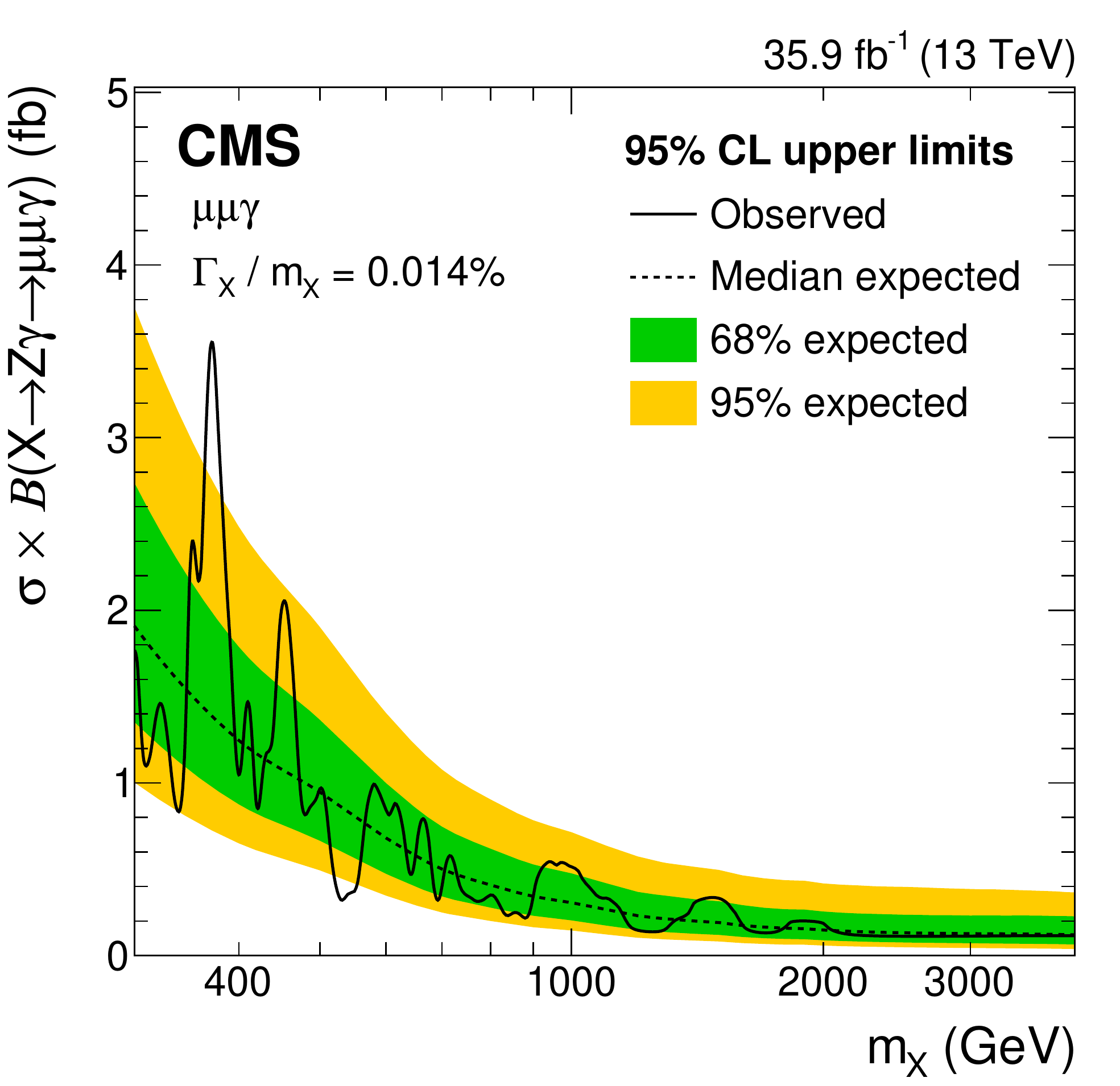}
      \includegraphics[width=0.48\textwidth]{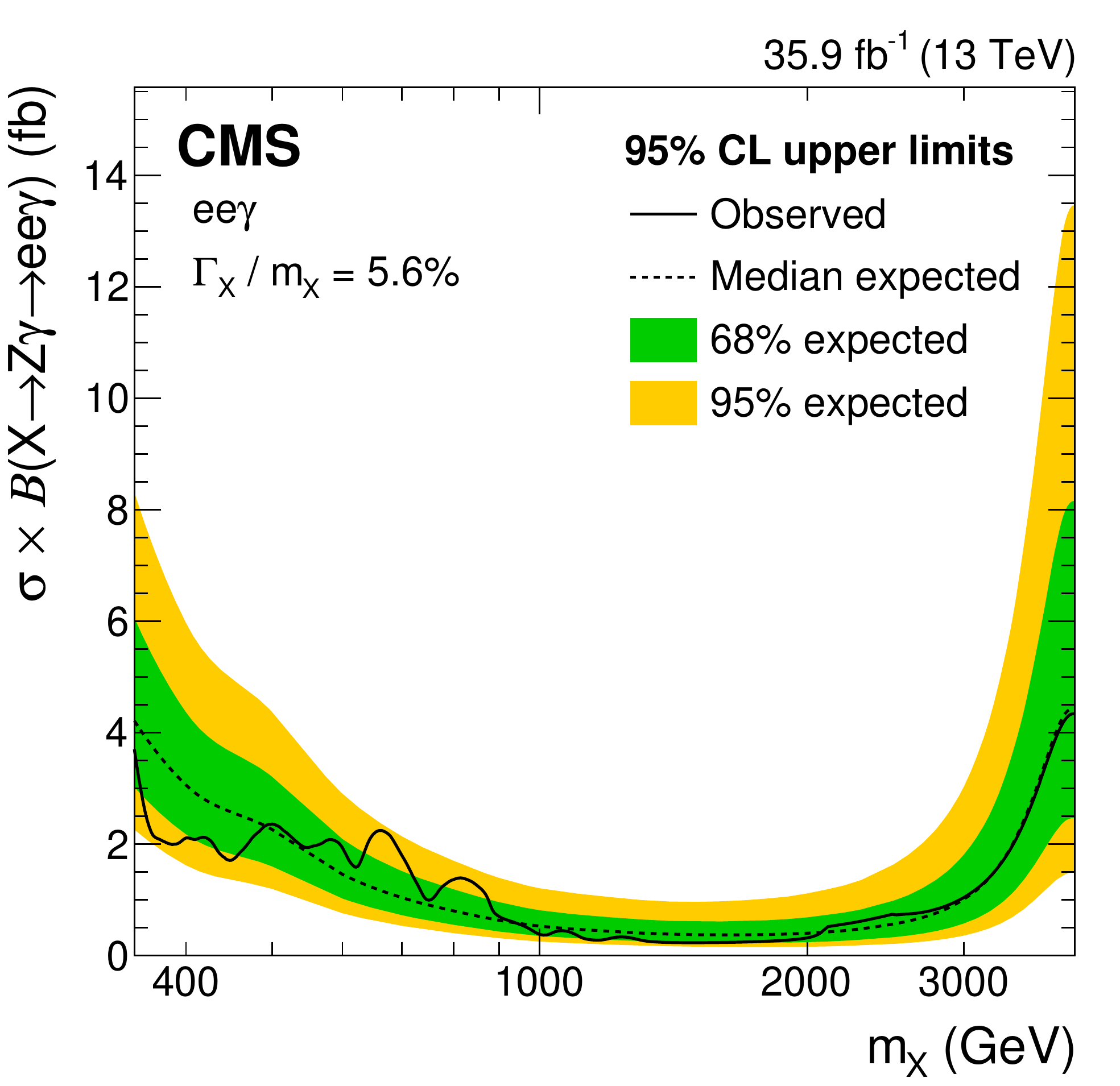}
      \includegraphics[width=0.48\textwidth]{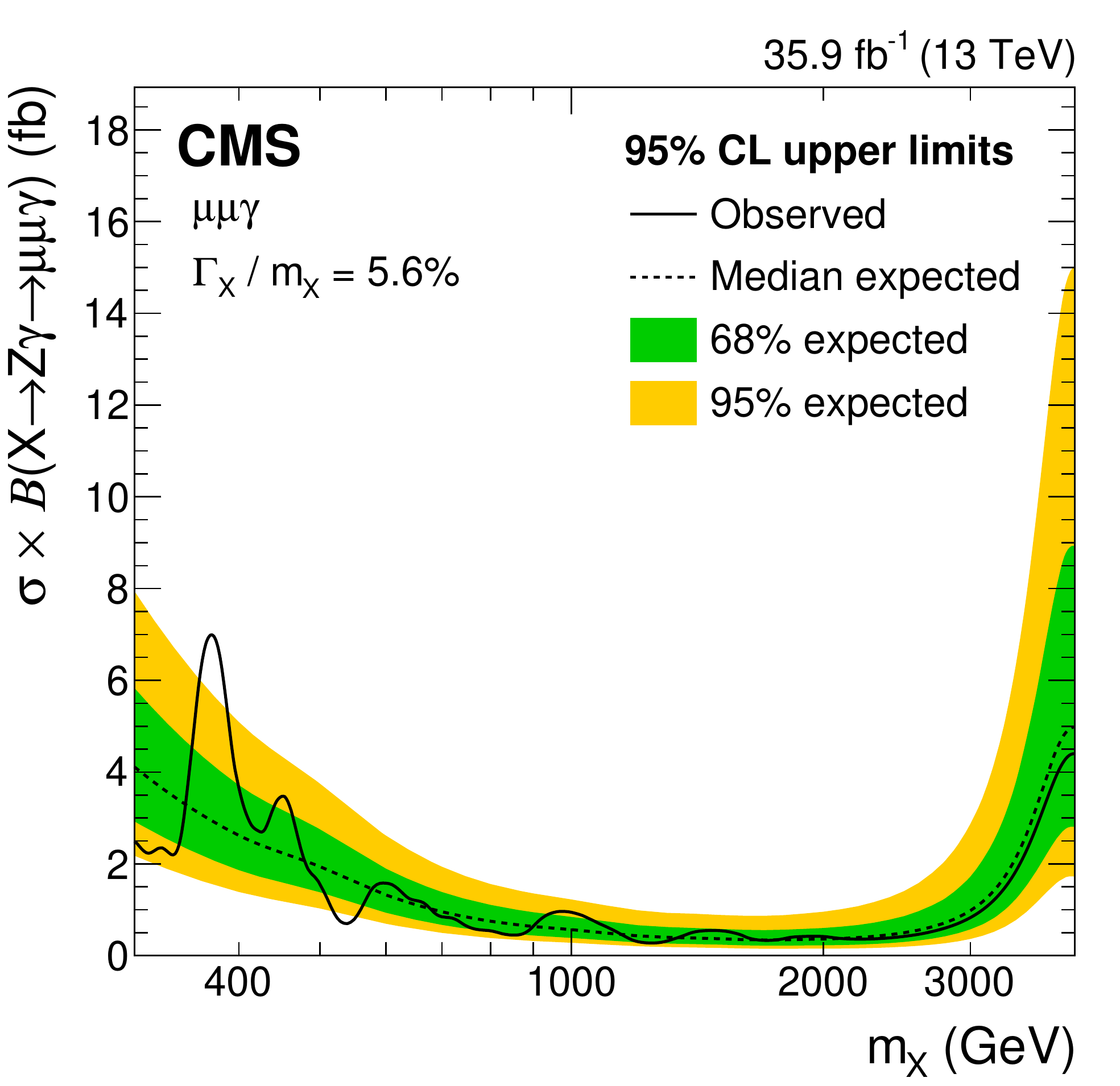}
    \caption{Observed~(solid) and expected~(dashed)  95\% CL upper limits on
 $\sigma(\mathrm{X}\to \PZ\Pgg)\,\mathcal{B}(\PZ \to \ell\ell\Pgg)$, as a function of signal
mass $m_\mathrm{X}$ for the \eeg (left column) and \mmg (right column) channels, and for narrow (upper row) and broad (lower row) spin-0 resonances. The green and yellow shaded bands correspond to respective 68 and 95\% CL ranges in the expected limits for the background-only hypothesis.
    }
    \label{fig:ul_em}
\end{figure}
\begin{figure}[htb!]
  \centering
      \includegraphics[width=0.48\textwidth]{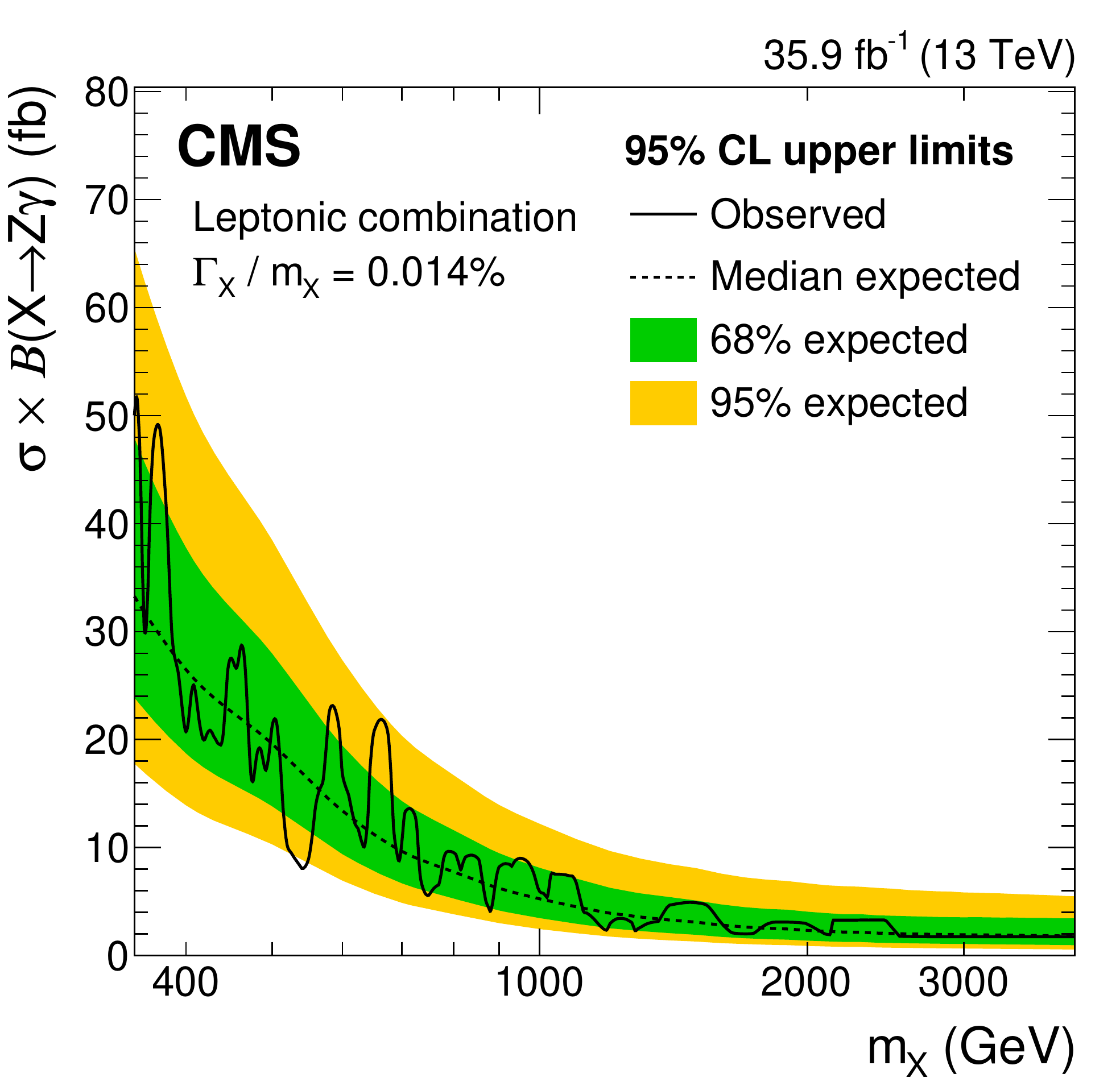}
      \includegraphics[width=0.48\textwidth]{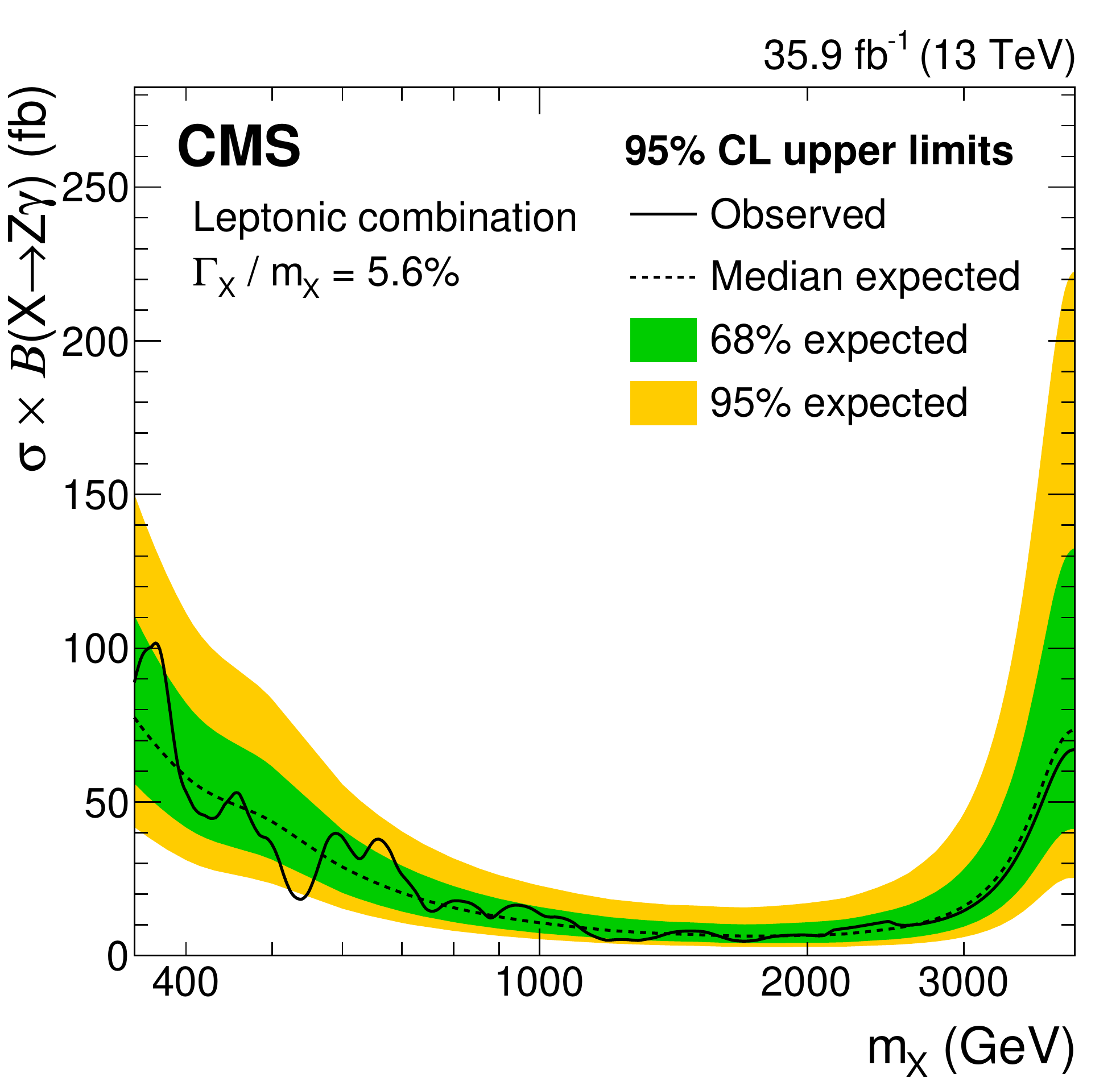}
    \caption{Observed~(solid) and expected~(dashed)  95\% CL upper limits
on  $\sigma(\mathrm{X}\to \PZ\Pgg)$ as a function of signal mass $m_\mathrm{X}$,
together with the 68\%~(green) and 95\%~(yellow) CL ranges of the expected limit for the
background-only hypothesis, for the combination of the \eeg and \mmg channels for (left) narrow and (right) broad spin-0 resonances.
    }
    \label{fig:ul}
\end{figure}

Figure~\ref{fig:ul_em} shows the observed and expected 95\% CL upper limits
on the product of signal cross section and branching fraction to the \llg final state, $\sigma(\mathrm{X}\to \PZ\Pgg)\,\mathcal{B}(\PZ \to \ell\ell\Pgg)$, as a function of the resonance mass,
for the \eeg channel~(left column) and the \mmg channel~(right column), for narrow (upper row) and broad (lower row) resonances.
The expected limits for the background-only hypothesis
are represented by the dashed black lines, and their 68 and 95\% CL ranges
are shown with the green and yellow bands, respectively. The observed limits
are represented by the solid black lines. The highest observed deviation is found in the \mmg channel at the mass of approximately 350\GeV and corresponds to a local (global) significance of approximately 3.0 (2.1) standard deviations for a narrow resonance. The limits on $\sigma(\mathrm{X}\to \PZ\Pgg)$, obtained by combining
the two leptonic search channels and taking into account the leptonic branching fraction of the Z boson decays~\cite{PDG}, are shown in Fig.~\ref{fig:ul}. The rapid increase in the limit for a broad resonance with a mass above approximately 3\TeV is due to a significant low-mass tail in the resonance line-shape extending outside the truncation window, as discussed in Section~\ref{sec:signal}.

\subsection{The \texorpdfstring{\Jg}{J gamma} channel}

\begin{figure}[htb!]
\centering
\includegraphics[width=0.32\textwidth]{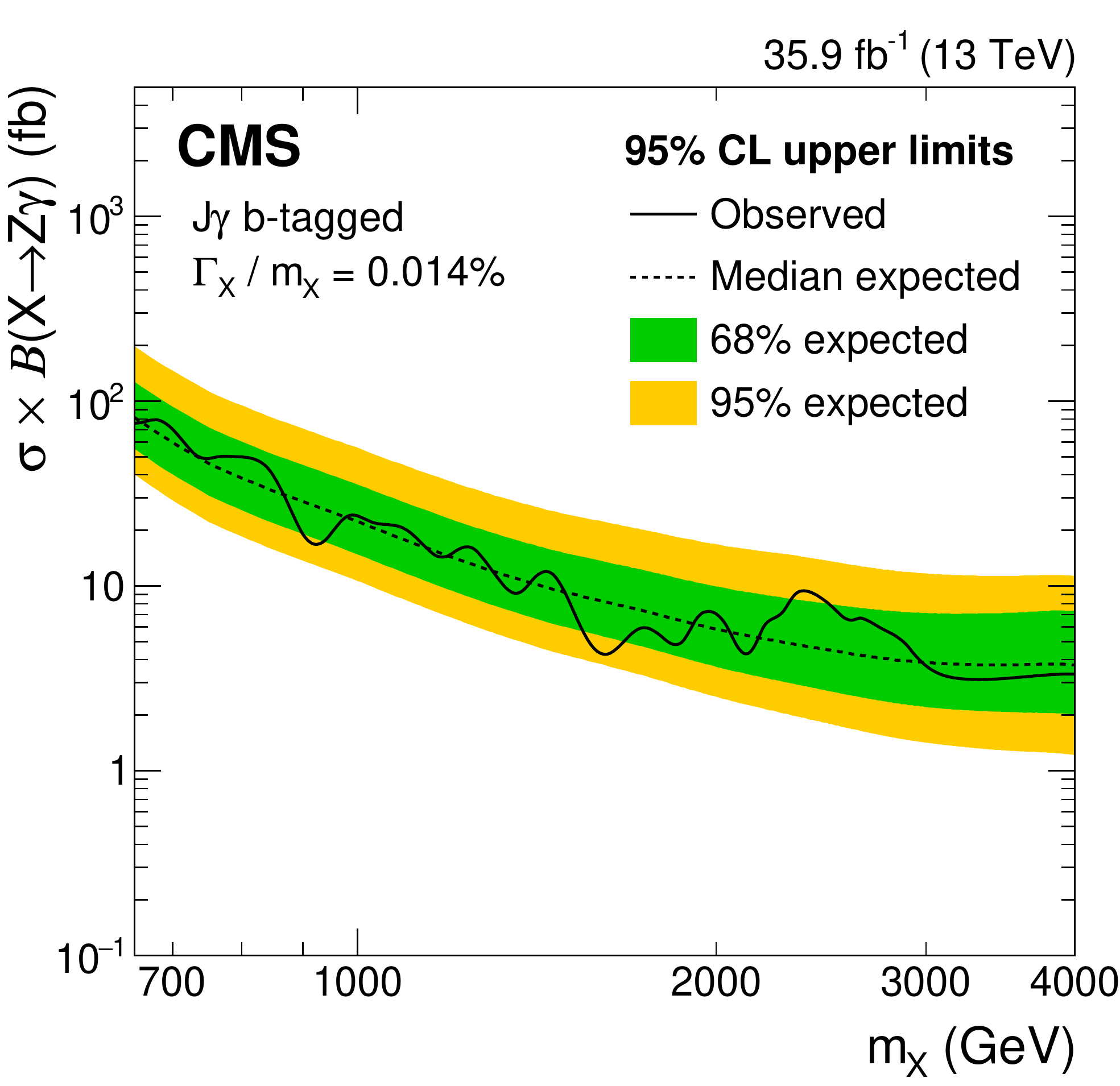}
\includegraphics[width=0.32\textwidth]{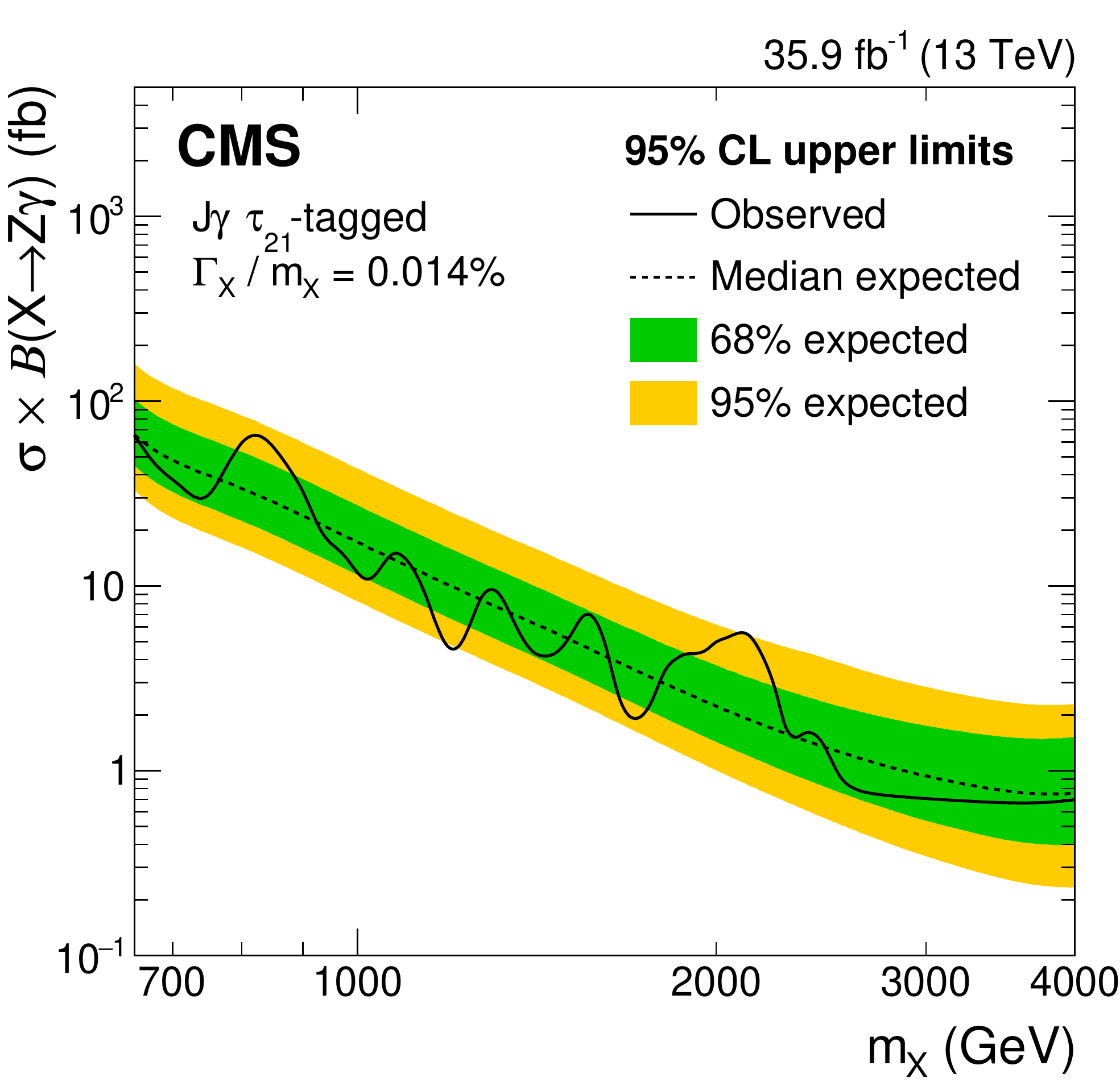}
\includegraphics[width=0.32\textwidth]{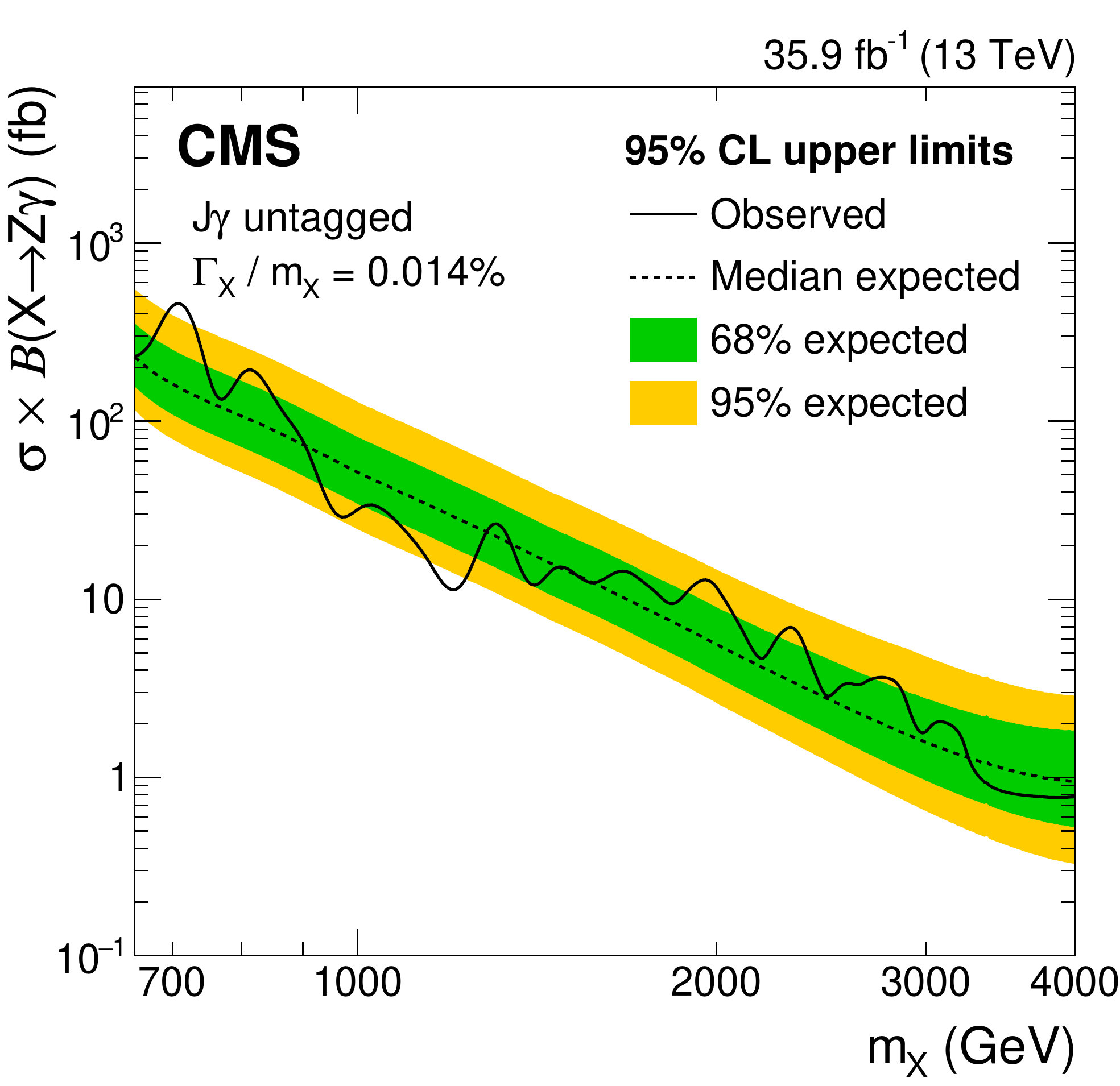}
\includegraphics[width=0.32\textwidth]{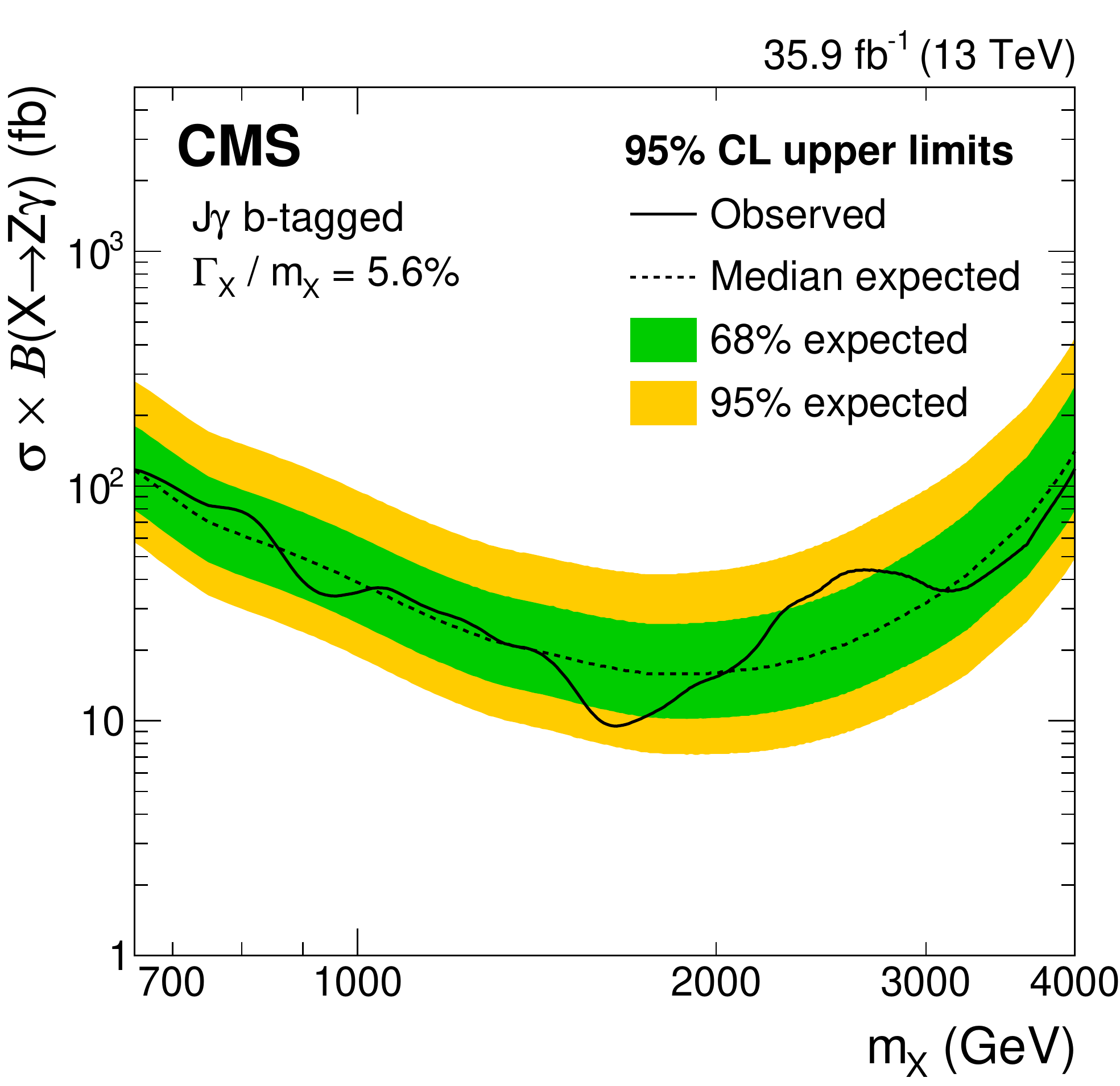}
\includegraphics[width=0.32\textwidth]{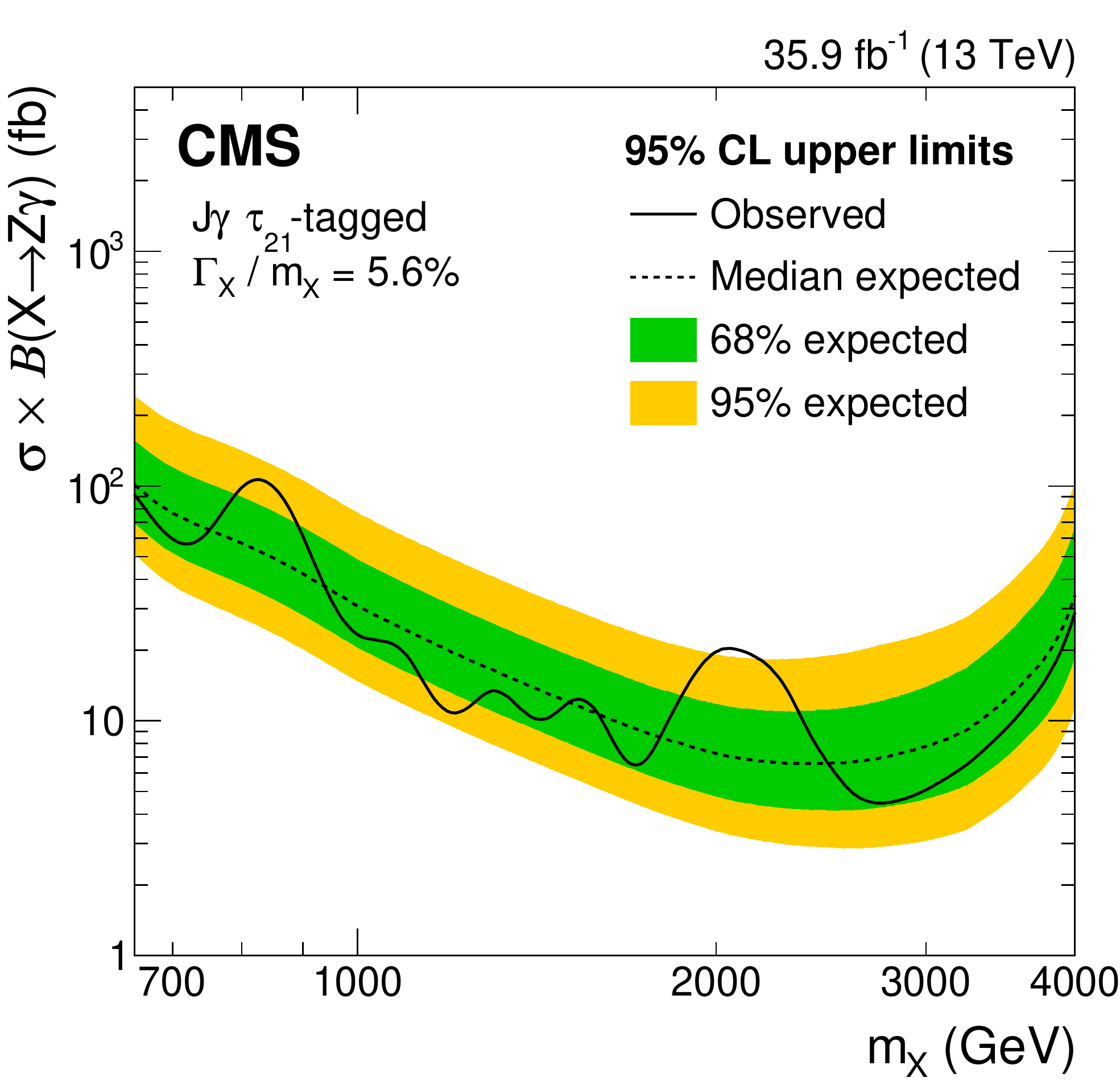}
\includegraphics[width=0.32\textwidth]{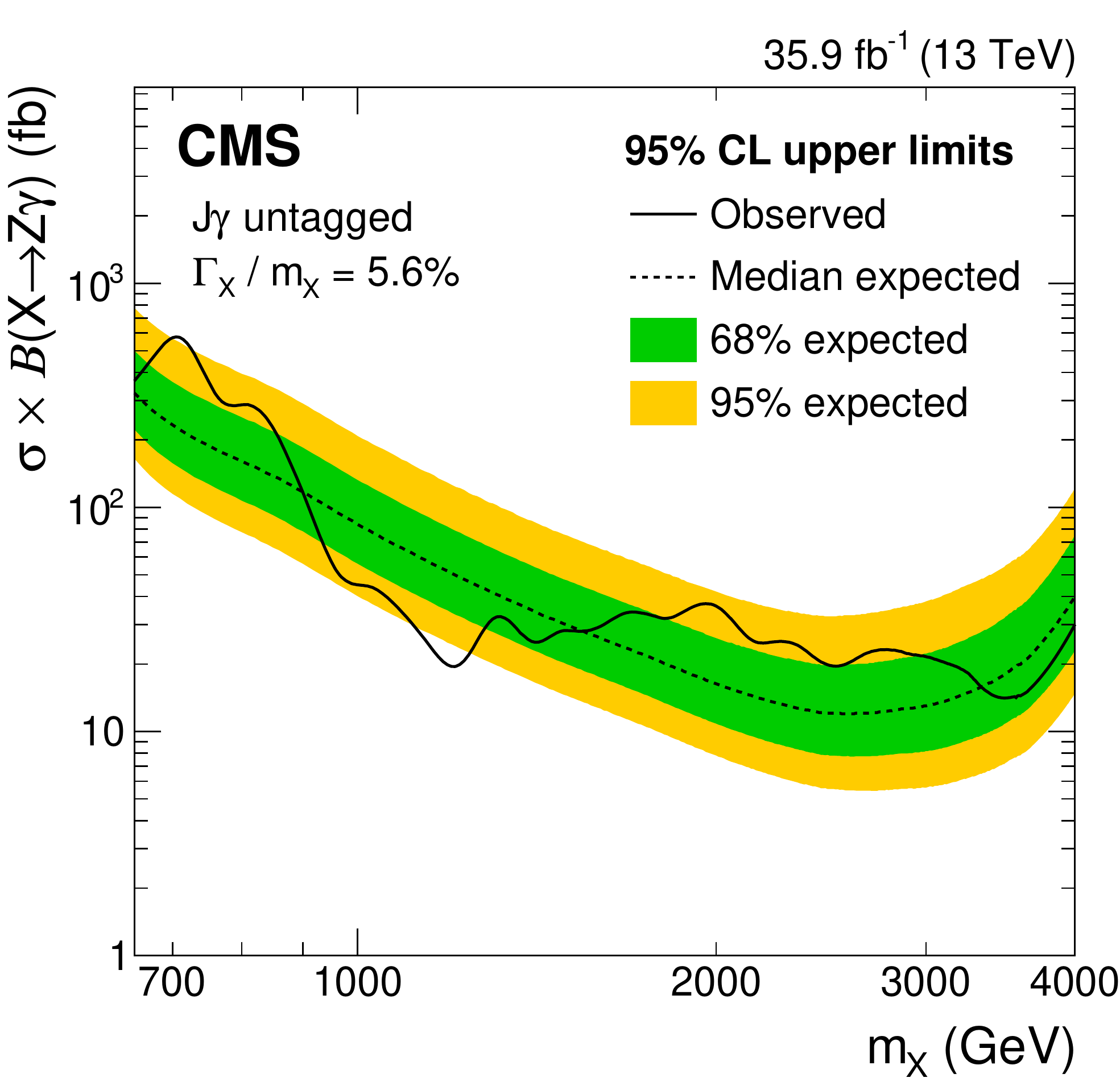}
\caption{Observed~(solid) and expected~(dashed)  95\% CL upper limits on
$\sigma(\mathrm{X}\to \PZ\Pgg)$, as a function of signal
mass $m_\mathrm{X}$, for the  b-tagged (left column), $\tau_{21}$-tagged (middle column), and untagged (right column) categories, and for narrow (upper row) and broad (lower row) spin-0 resonances. The colored bands correspond to the 68\% (green) and 95\% (yellow) CL ranges of the expected limit for the
background-only hypothesis.
}
\label{fig:limits}
\end{figure}

\begin{figure}[htb!]
\centering
\includegraphics[width=0.48\textwidth]{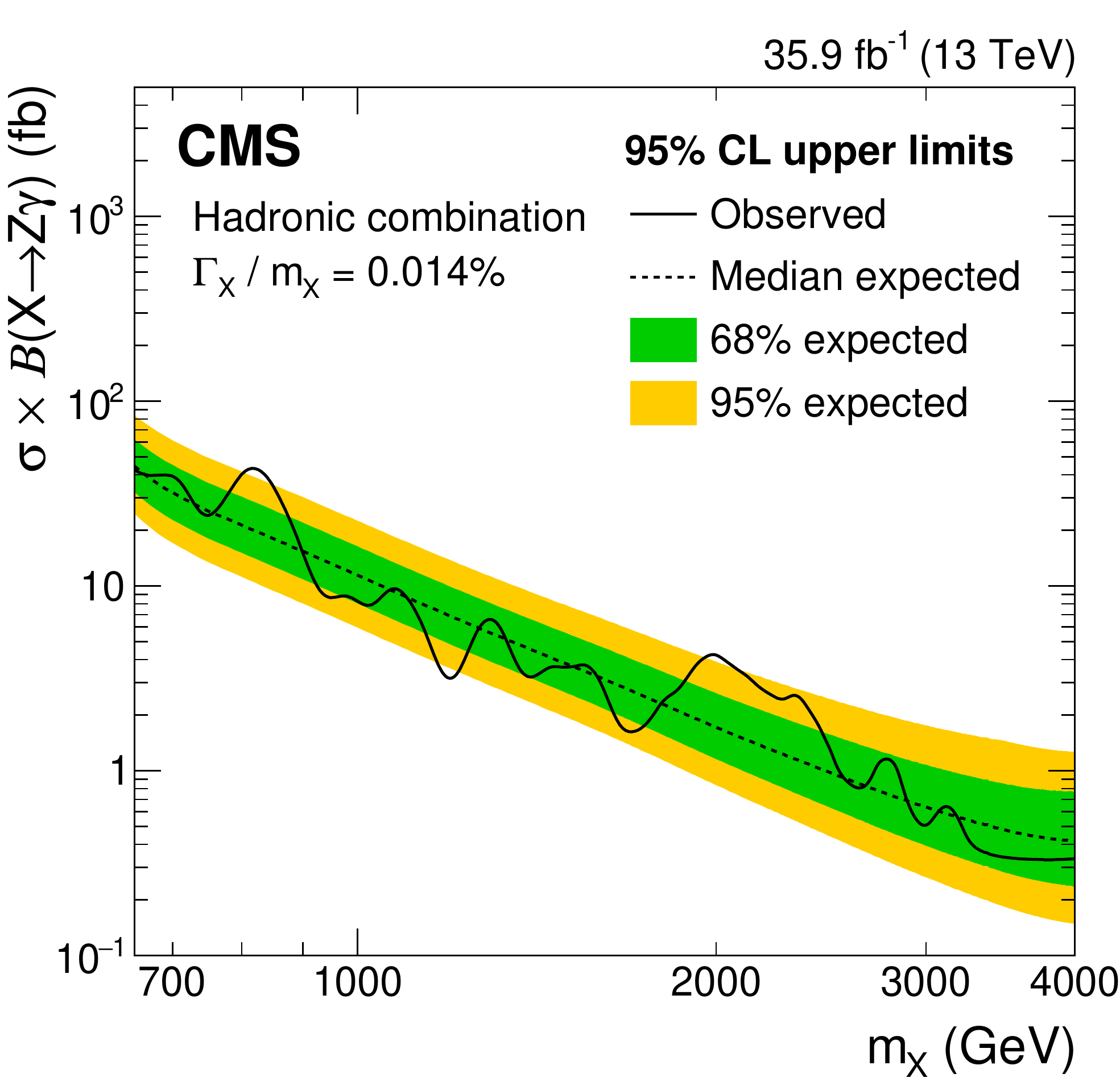}
\includegraphics[width=0.48\textwidth]{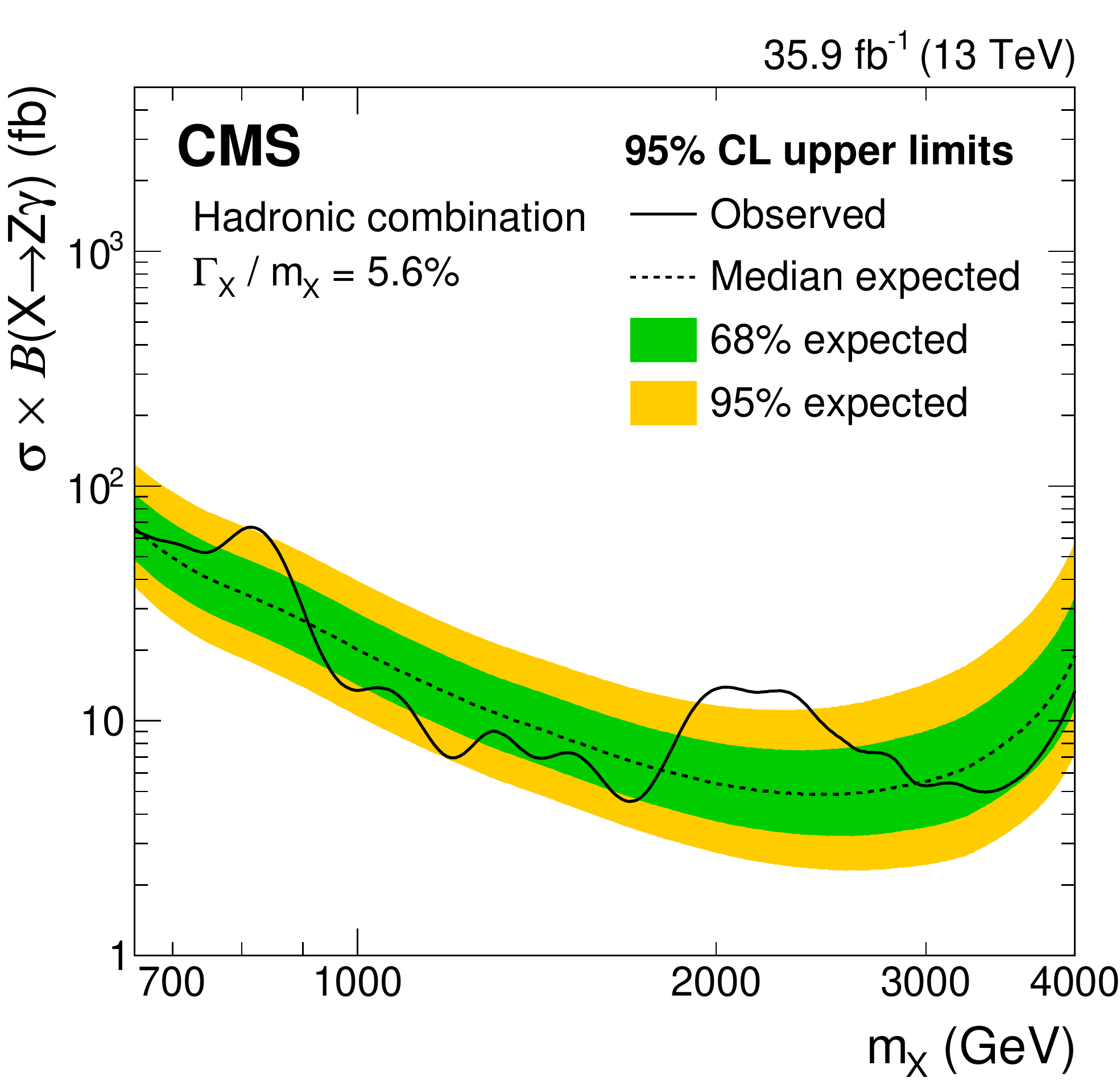}
\caption{Observed~(solid) and expected~(dashed)  95\% CL upper limits
on $\sigma(\mathrm{X}\to \PZ\Pgg)$ as a function of signal mass $m_\mathrm{X}$,
together with the 68\%~(green) and 95\%~(yellow) CL ranges of the expected limit for the
background-only hypothesis, for the combination of the b-tagged, $\tau_{21}$-tagged, and untagged categories for (left) narrow and (right) broad spin-0 resonances.}
\label{fig:combinedlimits}
\end{figure}

The observed and  expected 95\% CL upper limits on the product of signal cross section and branching fraction in the $\PZ\Pgg$ channel, $\sigma(\mathrm{X}\to \PZ\Pgg)$ for narrow and broad resonances in the b-tagged, $\tau_{21}$-tagged, and untagged categories
are presented in Fig.~\ref{fig:limits}.
The results based on the combination of the three categories for both narrow and broad resonances are shown in Fig.~\ref{fig:combinedlimits}. The combination includes the (anti)correlation of systematic uncertainties between the three categories. We observe a small deviation at a mass $\approx$2\TeV with local significance of 2.7\,(3.6) standard deviations for the narrow (broad) resonance width hypothesis. The global significance of this excess is 1.8\,(2.8) standard deviations.

\begin{figure}[htb]
\centering
\includegraphics[width=0.48\textwidth]{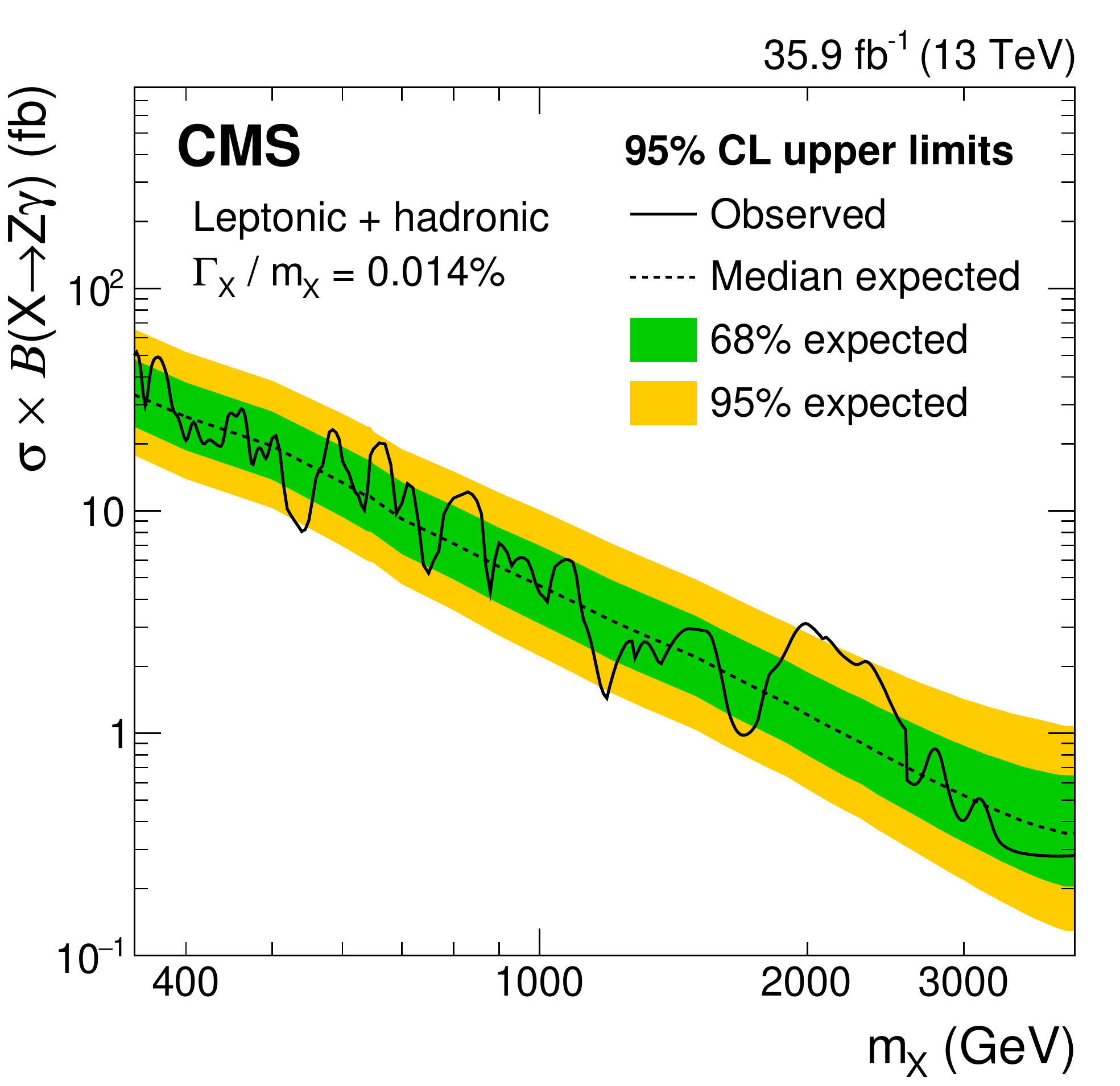}
\includegraphics[width=0.48\textwidth]{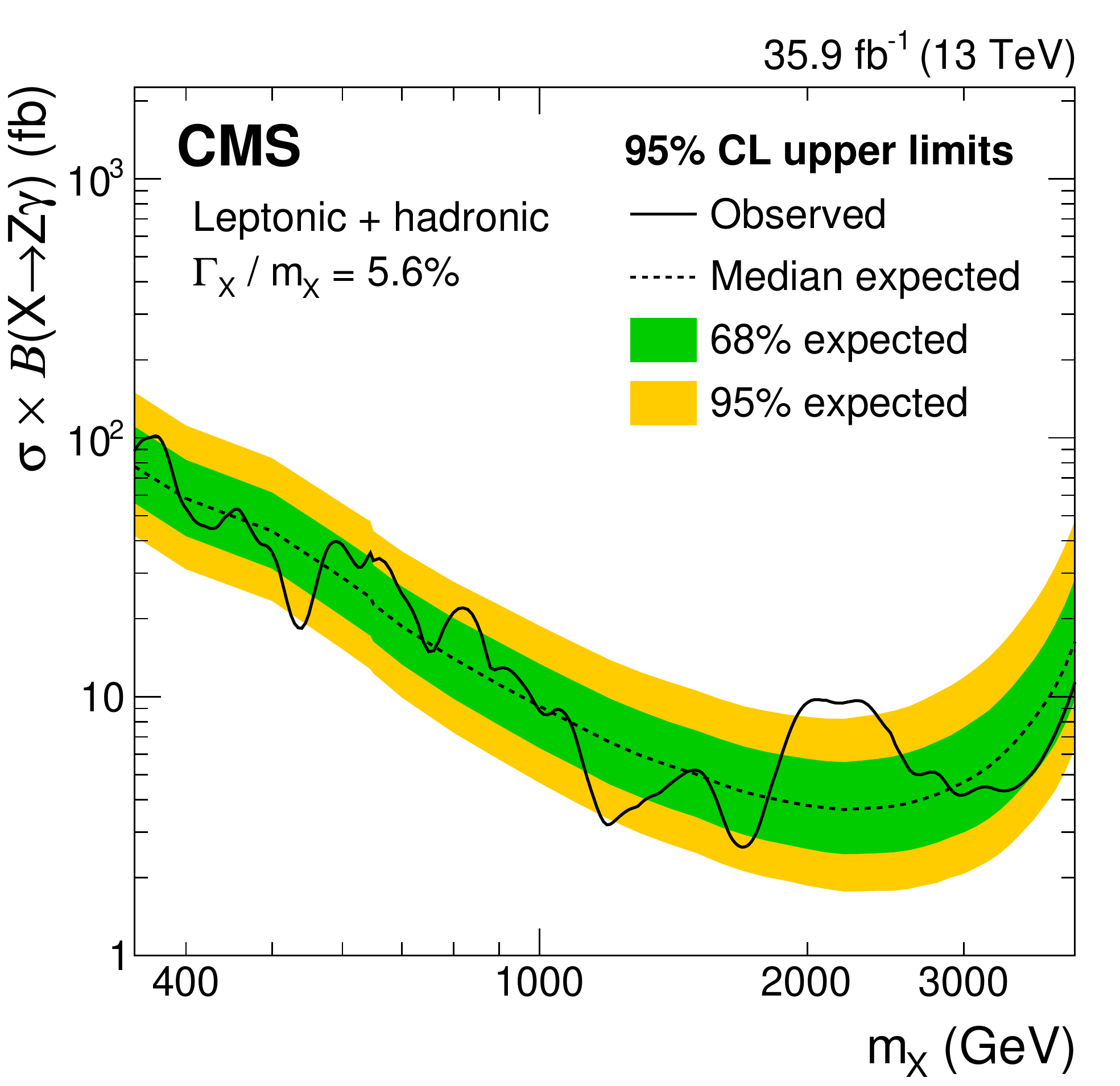}
\caption{Observed and expected limits on the product of the production cross section and branching fraction $\mathcal{B}(\mathrm{X} \to \PZ\Pgg)$, as a function of signal
mass $m_\mathrm{X}$,  for (left) narrow and (right) broad spin-0 resonances, obtained from the combination of the leptonic and hadronic decay channels.}
\label{fig:limitscomb}
\end{figure}

\subsection{Combination of the \texorpdfstring{\llg and \Jg}{ll gamma and J gamma} channels}

The results based on the combination of the \llg and \Jg channels are shown in Fig.~\ref{fig:limitscomb}, assuming uncorrelated uncertainties between the leptonic and hadronic channels, except for the uncertainties in the integrated luminosity, PDFs, and photon energy scale, which are taken as fully correlated among all the channels.
 These are the most stringent limits on resonances decaying in the $\PZ\Pgg$ channel to date in the mass range probed.
\section{Summary}
A search is presented for resonances decaying to a $\PZ$ boson and a photon. The analysis is based on data from proton-proton collisions at a center-of-mass energy of 13\TeV, corresponding to an integrated luminosity of 35.9\fbinv, collected with the CMS detector at the LHC in 2016. Two decay modes of the $\PZ$ boson are investigated. In the leptonic channels, the Z boson candidates are reconstructed using electron or muon pairs. In the hadronic channels, they are identified using a large-radius jet, containing either light-quark or b quark decay products of the $\PZ$ boson, via jet substructure and advanced b tagging techniques. The results from these channels are combined and interpreted in terms of upper limits on the product of the production cross section and the branching fraction to $\PZ\gamma$ for narrow (broad) spin-0 resonances with masses between 0.35 and 4.0\TeV, ranging from 50\,(100) to 0.3\,(1.5)\unit{fb}. These are the most stringent limits on such resonances to date.

\begin{acknowledgments}
We congratulate our colleagues in the CERN accelerator departments for the excellent performance of the LHC and thank the technical and administrative staffs at CERN and at other CMS institutes for their contributions to the success of the CMS effort. In addition, we gratefully acknowledge the computing centers and personnel of the Worldwide LHC Computing Grid for delivering so effectively the computing infrastructure essential to our analyses. Finally, we acknowledge the enduring support for the construction and operation of the LHC and the CMS detector provided by the following funding agencies: BMWFW and FWF (Austria); FNRS and FWO (Belgium); CNPq, CAPES, FAPERJ, and FAPESP (Brazil); MES (Bulgaria); CERN; CAS, MoST, and NSFC (China); COLCIENCIAS (Colombia); MSES and CSF (Croatia); RPF (Cyprus); SENESCYT (Ecuador); MoER, ERC IUT, and ERDF (Estonia); Academy of Finland, MEC, and HIP (Finland); CEA and CNRS/IN2P3 (France); BMBF, DFG, and HGF (Germany); GSRT (Greece); OTKA and NIH (Hungary); DAE and DST (India); IPM (Iran); SFI (Ireland); INFN (Italy); MSIP and NRF (Republic of Korea); LAS (Lithuania); MOE and UM (Malaysia); BUAP, CINVESTAV, CONACYT, LNS, SEP, and UASLP-FAI (Mexico); MBIE (New Zealand); PAEC (Pakistan); MSHE and NSC (Poland); FCT (Portugal); JINR (Dubna); MON, RosAtom, RAS, RFBR and RAEP (Russia); MESTD (Serbia); SEIDI, CPAN, PCTI and FEDER (Spain); Swiss Funding Agencies (Switzerland); MST (Taipei); ThEPCenter, IPST, STAR, and NSTDA (Thailand); TUBITAK and TAEK (Turkey); NASU and SFFR (Ukraine); STFC (United Kingdom); DOE and NSF (USA).

\hyphenation{Rachada-pisek} Individuals have received support from the Marie-Curie program and the European Research Council and Horizon 2020 Grant, contract No. 675440 (European Union); the Leventis Foundation; the A. P. Sloan Foundation; the Alexander von Humboldt Foundation; the Belgian Federal Science Policy Office; the Fonds pour la Formation \`a la Recherche dans l'Industrie et dans l'Agriculture (FRIA-Belgium); the Agentschap voor Innovatie door Wetenschap en Technologie (IWT-Belgium); the Ministry of Education, Youth and Sports (MEYS) of the Czech Republic; the Council of Science and Industrial Research, India; the HOMING PLUS program of the Foundation for Polish Science, cofinanced from European Union, Regional Development Fund, the Mobility Plus program of the Ministry of Science and Higher Education, the National Science Center (Poland), contracts Harmonia 2014/14/M/ST2/00428, Opus 2014/13/B/ST2/02543, 2014/15/B/ST2/03998, and 2015/19/B/ST2/02861, Sonata-bis 2012/07/E/ST2/01406; the National Priorities Research Program by Qatar National Research Fund; the Programa Severo Ochoa del Principado de Asturias; the Thalis and Aristeia programs cofinanced by EU-ESF and the Greek NSRF; the Rachadapisek Sompot Fund for Postdoctoral Fellowship, Chulalongkorn University and the Chulalongkorn Academic into Its 2nd Century Project Advancement Project (Thailand); the Welch Foundation, contract C-1845; and the Weston Havens Foundation (USA).
\end{acknowledgments}

\bibliography{auto_generated}

\providecommand{\href}[2]{#2}\begingroup\raggedright\begin{thebibliography}{10}%
\makeatletter
\providecommand{\hrefCMSnoop }[0]{\@secondoftwo}%
\makeatother
\providecommand{\doi}{\texttt{doi:}\begingroup \urlstyle{tt}\Url}

\bibitem{Higgs1}
\hrefCMSnoop {}{{ATLAS Collaboration}, ``{Observation of a new particle in the
  search for the Standard Model Higgs boson with the ATLAS detector at the
  LHC}'',} \textit{ Phys. Lett. B} \textbf{ 716} (2012) 1,
  \href{http://dx.doi.org/10.1016/j.physletb.2012.08.020}{\doi{10.1016/j.physletb.2012.08.020}},
\href{http://www.arXiv.org/abs/1207.7214}{\texttt{arXiv:1207.7214}}.

\bibitem{Higgs2}
\hrefCMSnoop {}{{CMS Collaboration}, ``{Observation of a new boson at a mass of
  125 GeV with the CMS experiment at the LHC}'',} \textit{ Phys. Lett. B}
  \textbf{ 716} (2012) 30,
  \href{http://dx.doi.org/10.1016/j.physletb.2012.08.021}{\doi{10.1016/j.physletb.2012.08.021}},
\href{http://www.arXiv.org/abs/1207.7235}{\texttt{arXiv:1207.7235}}.

\bibitem{Higgs3}
\hrefCMSnoop {}{{CMS Collaboration}, ``{Observation of a new boson with mass
  near 125 GeV in pp collisions at $\sqrt{s} = 7$ and 8 TeV}'',} \textit{ JHEP}
  \textbf{ 06} (2013) 081,
  \href{http://dx.doi.org/10.1007/JHEP06(2013)081}{\doi{10.1007/JHEP06(2013)081}},
\href{http://www.arXiv.org/abs/1303.4571}{\texttt{arXiv:1303.4571}}.

\bibitem{diboson1}
\hrefCMSnoop {}{{CMS Collaboration}, ``{Search for massive resonances in dijet
  systems containing jets tagged as W or Z boson decays in pp collisions at $
  \sqrt{s} = 8$ TeV}'',} \textit{ JHEP} \textbf{ 08} (2014) 173,
  \href{http://dx.doi.org/10.1007/JHEP08(2014)173}{\doi{10.1007/JHEP08(2014)173}},
\href{http://www.arXiv.org/abs/1405.1994}{\texttt{arXiv:1405.1994}}.

\bibitem{diboson2}
\hrefCMSnoop {}{{CMS Collaboration}, ``{Search for massive resonances decaying
  into pairs of boosted bosons in semi-leptonic final states at $\sqrt{s} = 8$
  TeV}'',} \textit{ JHEP} \textbf{ 08} (2014) 174,
  \href{http://dx.doi.org/10.1007/JHEP08(2014)174}{\doi{10.1007/JHEP08(2014)174}},
\href{http://www.arXiv.org/abs/1405.3447}{\texttt{arXiv:1405.3447}}.

\bibitem{diboson3}
\hrefCMSnoop {}{{ATLAS Collaboration}, ``{Combination of searches for WW, WZ,
  and ZZ resonances in pp collisions at $\sqrt{s} = 8$ TeV with the ATLAS
  detector}'',} \textit{ Phys. Lett. B} \textbf{ 755} (2016) 285,
  \href{http://dx.doi.org/10.1016/j.physletb.2016.02.015}{\doi{10.1016/j.physletb.2016.02.015}},
\href{http://www.arXiv.org/abs/1512.05099}{\texttt{arXiv:1512.05099}}.

\bibitem{diboson4}
\hrefCMSnoop {}{{ATLAS Collaboration}, ``{Searches for heavy diboson resonances
  in pp collisions at $\sqrt{s}=13$ TeV with the ATLAS detector}'',} \textit{
  JHEP} \textbf{ 09} (2016) 173,
  \href{http://dx.doi.org/10.1007/JHEP09(2016)173}{\doi{10.1007/JHEP09(2016)173}},
\href{http://www.arXiv.org/abs/1606.04833}{\texttt{arXiv:1606.04833}}.

\bibitem{diboson5}
\hrefCMSnoop {}{{CMS Collaboration}, ``{Search for massive resonances decaying
  into WW, WZ or ZZ bosons in proton-proton collisions at $\sqrt{s} = $ 13
  TeV}'',} \textit{ JHEP} \textbf{ 03} (2017) 162,
  \href{http://dx.doi.org/10.1007/JHEP03(2017)162}{\doi{10.1007/JHEP03(2017)162}},
\href{http://www.arXiv.org/abs/1612.09159}{\texttt{arXiv:1612.09159}}.

\bibitem{diboson6}
\hrefCMSnoop {}{{CMS Collaboration}, ``Search for charged higgs bosons produced
  via vector boson fusion and decaying into a pair of {$W$} and {$Z$} bosons
  using $pp$ collisions at {$\sqrt{s}=13$ TeV}'',} \textit{ Phys. Rev. Lett.}
  \textbf{ 119} (2017) 141802,
  \href{http://dx.doi.org/10.1103/PhysRevLett.119.141802}{\doi{10.1103/PhysRevLett.119.141802}},
\href{http://www.arXiv.org/abs/1705.02942}{\texttt{arXiv:1705.02942}}.

\bibitem{diboson7}
\hrefCMSnoop {}{{CMS Collaboration}, ``{Combination of searches for heavy
  resonances decaying to WW, WZ, ZZ, WH, and ZH boson pairs in proton?proton
  collisions at $\sqrt{s}=8$ and 13 TeV}'',} \textit{ Phys. Lett. B} \textbf{
  774} (2017) 533,
  \href{http://dx.doi.org/10.1016/j.physletb.2017.09.083}{\doi{10.1016/j.physletb.2017.09.083}},
\href{http://www.arXiv.org/abs/1705.09171}{\texttt{arXiv:1705.09171}}.

\bibitem{diboson8}
\hrefCMSnoop {}{{ATLAS Collaboration}, ``Search for diboson resonances with
  boson-tagged jets in $pp$ collisions at $\sqrt{s}=13$ {TeV} with the {ATLAS}
  detector'',} \textit{ Phys. Lett. B} \textbf{ 777} (2018) 91,
  \href{http://dx.doi.org/10.1016/j.physletb.2017.12.011}{\doi{10.1016/j.physletb.2017.12.011}},
\href{http://www.arXiv.org/abs/1708.04445}{\texttt{arXiv:1708.04445}}.

\bibitem{diboson9}
\hrefCMSnoop {}{{ATLAS Collaboration}, ``Searches for heavy {$ZZ$} and {$ZW$}
  resonances in the $\ell\ell qq$ and $\nu\nu qq$ final states in $pp$
  collisions at $\sqrt{s}=13$ {TeV} with the {ATLAS} detector'',} \textit{
  JHEP} \textbf{ 03} (2018) 009,
  \href{http://dx.doi.org/10.1007/JHEP03(2018)009}{\doi{10.1007/JHEP03(2018)009}},
\href{http://www.arXiv.org/abs/1708.09638}{\texttt{arXiv:1708.09638}}.

\bibitem{diboson10}
\hrefCMSnoop {}{{ATLAS Collaboration}, ``Search for {$WW/WZ$} resonance
  production in $\ell \nu qq$ final states in $pp$ collisions at $\sqrt{s} =$
  13 {TeV} with the {ATLAS} detector'',} \textit{ JHEP} \textbf{ 03} (2018)
  042,
  \href{http://dx.doi.org/10.1007/JHEP03(2018)042}{\doi{10.1007/JHEP03(2018)042}},
\href{http://www.arXiv.org/abs/1710.07235}{\texttt{arXiv:1710.07235}}.

\bibitem{diboson11}
\hrefCMSnoop {}{{ATLAS Collaboration}, ``Search for heavy resonances decaying
  into {$WW$} in the $e\nu\mu\nu$ final state in $pp$ collisions at
  $\sqrt{s}=13$ {TeV} with the {ATLAS} detector'',} \textit{ Eur. Phys. J. C}
  \textbf{ 78} (2018) 24,
  \href{http://dx.doi.org/10.1140/epjc/s10052-017-5491-4}{\doi{10.1140/epjc/s10052-017-5491-4}},
\href{http://www.arXiv.org/abs/1710.01123}{\texttt{arXiv:1710.01123}}.

\bibitem{diboson12}
\hrefCMSnoop {}{{CMS Collaboration}, ``Search for {ZZ} resonances in the
  2$\ell$2$\nu$ final state in proton-proton collisions at 13 {TeV}'',}
  \textit{ JHEP} \textbf{ 03} (2018) 003,
  \href{http://dx.doi.org/10.1007/JHEP03(2018)003}{\doi{10.1007/JHEP03(2018)003}},
\href{http://www.arXiv.org/abs/1711.04370}{\texttt{arXiv:1711.04370}}.

\bibitem{VH1}
\hrefCMSnoop {}{{CMS Collaboration}, ``{Search for a massive resonance decaying
  into a Higgs boson and a W or Z boson in hadronic final states in
  proton-proton collisions at $ \sqrt{s}=8 $ TeV}'',} \textit{ JHEP} \textbf{
  02} (2016) 145,
  \href{http://dx.doi.org/10.1007/JHEP02(2016)145}{\doi{10.1007/JHEP02(2016)145}},
\href{http://www.arXiv.org/abs/1506.01443}{\texttt{arXiv:1506.01443}}.

\bibitem{VH2}
\hrefCMSnoop {}{{CMS Collaboration}, ``{Search for massive WH resonances
  decaying into the $\ell \nu \mathrm{b} \overline{\mathrm{b}} $ final state at
  $\sqrt{s}=8$ $~\text {TeV}$}'',} \textit{ Eur. Phys. J. C} \textbf{ 76}
  (2016) 237,
  \href{http://dx.doi.org/10.1140/epjc/s10052-016-4067-z}{\doi{10.1140/epjc/s10052-016-4067-z}},
\href{http://www.arXiv.org/abs/1601.06431}{\texttt{arXiv:1601.06431}}.

\bibitem{VH3}
\hrefCMSnoop {}{{ATLAS Collaboration}, ``{Search for new resonances decaying to
  a $W$ or $Z$ boson and a Higgs boson in the $\ell^+
  \ell^-\text{b}\bar{\text{b}}$, $\ell \nu\text{b}\bar{\text{b}}$, and
  $\nu\bar{\nu}\text{b}\bar{\text{b}}$ channels with pp collisions at $\sqrt s
  = 13$ TeV with the ATLAS detector}'',} \textit{ Phys. Lett. B} \textbf{ 765}
  (2017) 32,
  \href{http://dx.doi.org/10.1016/j.physletb.2016.11.045}{\doi{10.1016/j.physletb.2016.11.045}},
\href{http://www.arXiv.org/abs/1607.05621}{\texttt{arXiv:1607.05621}}.

\bibitem{VH4}
\hrefCMSnoop {}{{CMS Collaboration}, ``{Search for heavy resonances decaying
  into a vector boson and a Higgs boson in final states with charged leptons,
  neutrinos, and b quarks}'',} \textit{ Phys. Lett. B} \textbf{ 768} (2017)
  137,
  \href{http://dx.doi.org/10.1016/j.physletb.2017.02.040}{\doi{10.1016/j.physletb.2017.02.040}},
\href{http://www.arXiv.org/abs/1610.08066}{\texttt{arXiv:1610.08066}}.

\bibitem{VH5}
\hrefCMSnoop {}{{CMS Collaboration}, ``{Search for heavy resonances that decay
  into a vector boson and a Higgs boson in hadronic final states at $\sqrt{s} =
  13$ $\,\text {TeV}$}'',} \textit{ Eur. Phys. J. C} \textbf{ 77} (2017) 636,
  \href{http://dx.doi.org/10.1140/epjc/s10052-017-5192-z}{\doi{10.1140/epjc/s10052-017-5192-z}},
\href{http://www.arXiv.org/abs/1707.01303}{\texttt{arXiv:1707.01303}}.

\bibitem{VH6}
\hrefCMSnoop {}{{ATLAS Collaboration}, ``{Search for heavy resonances decaying
  to a $W$ or $Z$ boson and a Higgs boson in the $q\bar{q}^{(\prime)}b\bar{b}$
  final state in $pp$ collisions at $\sqrt{s} = 13$ TeV with the ATLAS
  detector}'',} \textit{ Phys. Lett. B} \textbf{ 774} (2017) 494,
  \href{http://dx.doi.org/10.1016/j.physletb.2017.09.066}{\doi{10.1016/j.physletb.2017.09.066}},
\href{http://www.arXiv.org/abs/1707.06958}{\texttt{arXiv:1707.06958}}.

\bibitem{Landau}
\hrefCMSnoop {}{L.~D. Landau, ``{On the angular momentum of a system of two
  photons}'',} \textit{ Dokl. Akad. Nauk Ser. Fiz.} \textbf{ 60} (1948) 207,
\href{http://dx.doi.org/10.1016/B978-0-08-010586-4.50070-5}{\doi{10.1016/B978-0-08-010586-4.50070-5}}.

\bibitem{Yang}
\hrefCMSnoop {}{C.-N. Yang, ``{Selection rules for the dematerialization of a
  particle into two photons}'',} \textit{ Phys. Rev.} \textbf{ 77} (1950) 242,
\href{http://dx.doi.org/10.1103/PhysRev.77.242}{\doi{10.1103/PhysRev.77.242}}.

\bibitem{EL}
\hrefCMSnoop {}{E.~Eichten and K.~Lane, ``{Low-scale technicolor at the
  Tevatron and LHC}'',} \textit{ Phys. Lett. B} \textbf{ 669} (2008) 235,
  \href{http://dx.doi.org/10.1016/j.physletb.2008.09.047}{\doi{10.1016/j.physletb.2008.09.047}},
\href{http://www.arXiv.org/abs/0706.2339}{\texttt{arXiv:0706.2339}}.

\bibitem{FS}
\hrefCMSnoop {}{A.~Freitas and P.~Schwaller, ``Multi-photon signals from
  composite models at {LHC}'',} \textit{ JHEP} \textbf{ 01} (2011) 022,
  \href{http://dx.doi.org/10.1007/JHEP01(2011)022}{\doi{10.1007/JHEP01(2011)022}},
\href{http://www.arXiv.org/abs/1010.2528}{\texttt{arXiv:1010.2528}}.

\bibitem{BT}
\hrefCMSnoop {}{R.~Barbieri and R.~Torre, ``{Signals of single particle
  production at the earliest LHC}'',} \textit{ Phys. Lett. B} \textbf{ 695}
  (2011) 259,
  \href{http://dx.doi.org/10.1016/j.physletb.2010.11.037}{\doi{10.1016/j.physletb.2010.11.037}},
\href{http://www.arXiv.org/abs/1008.5302}{\texttt{arXiv:1008.5302}}.

\bibitem{LLS}
\hrefCMSnoop {}{I.~Low, J.~Lykken, and G.~Shaughnessy, ``{Singlet scalars as
  Higgs imposters at the Large Hadron Collider}'',} \textit{ Phys. Rev. D}
  \textbf{ 84} (2011) 035027,
  \href{http://dx.doi.org/10.1103/PhysRevD.84.035027}{\doi{10.1103/PhysRevD.84.035027}},
\href{http://www.arXiv.org/abs/1105.4587}{\texttt{arXiv:1105.4587}}.

\bibitem{DHR}
\hrefCMSnoop {}{H.~Davoudiasl, J.~L. Hewett, and T.~G. Rizzo, ``{Experimental
  probes of localized gravity: On and off the wall}'',} \textit{ Phys. Rev. D}
  \textbf{ 63} (2001) 075004,
  \href{http://dx.doi.org/10.1103/PhysRevD.63.075004}{\doi{10.1103/PhysRevD.63.075004}},
\href{http://www.arXiv.org/abs/hep-ph/0006041}{\texttt{arXiv:hep-ph/0006041}}.

\bibitem{ASS}
\hrefCMSnoop {}{B.~C. Allanach, J.~P. Skittrall, and K.~Sridhar, ``{Z boson
  decay to photon plus Kaluza-Klein graviton in large extra dimensions}'',}
  \textit{ JHEP} \textbf{ 11} (2007) 089,
  \href{http://dx.doi.org/10.1088/1126-6708/2007/11/089}{\doi{10.1088/1126-6708/2007/11/089}},
\href{http://www.arXiv.org/abs/0705.1953}{\texttt{arXiv:0705.1953}}.

\bibitem{Aaboud:2017uhw}
\hrefCMSnoop {}{{ATLAS Collaboration}, ``{Searches for the $Z\gamma$ decay mode
  of the Higgs boson and for new high-mass resonances in $pp$ collisions at
  $\sqrt{s} = 13$ TeV with the ATLAS detector}'',} \textit{ JHEP} \textbf{ 10}
  (2017) 112,
  \href{http://dx.doi.org/10.1007/JHEP10(2017)112}{\doi{10.1007/JHEP10(2017)112}},
\href{http://www.arXiv.org/abs/1708.00212}{\texttt{arXiv:1708.00212}}.

\bibitem{ATLAS-Vgamma3}
\hrefCMSnoop {}{{ATLAS Collaboration}, ``{Search for heavy resonances decaying
  to a \PZ boson and a photon in pp collisions at $\sqrt{s}=13$ TeV with the
  ATLAS detector}'',} \textit{ Phys. Lett. B} \textbf{ 764} (2017) 11,
  \href{http://dx.doi.org/10.1016/j.physletb.2016.11.005}{\doi{10.1016/j.physletb.2016.11.005}},
\href{http://www.arXiv.org/abs/1607.06363}{\texttt{arXiv:1607.06363}}.

\bibitem{Zgamma-had}
\hrefCMSnoop {}{{CMS Collaboration}, ``{Search for high-mass $\mathrm{ Z
  }\gamma$ resonances in proton-proton collisions at $\sqrt{s}=$ 8 and 13 TeV
  using jet substructure techniques}'',} \textit{ Phys. Lett. B} \textbf{ 772}
  (2017) 363,
  \href{http://dx.doi.org/10.1016/j.physletb.2017.06.062}{\doi{10.1016/j.physletb.2017.06.062}},
\href{http://www.arXiv.org/abs/1612.09516}{\texttt{arXiv:1612.09516}}.

\bibitem{L3}
\hrefCMSnoop {}{{L3} Collaboration, ``{Search for anomalous couplings in the
  Higgs sector at LEP}'',} \textit{ Phys. Lett. B} \textbf{ 589} (2004) 89,
  \href{http://dx.doi.org/10.1016/j.physletb.2004.03.048}{\doi{10.1016/j.physletb.2004.03.048}},
\href{http://www.arXiv.org/abs/hep-ex/0403037}{\texttt{arXiv:hep-ex/0403037}}.

\bibitem{D0-1}
\hrefCMSnoop {}{{D0} Collaboration, ``{Search for particles decaying into a Z
  boson and a photon in ${\rm p\bar{p}}$ collisions at $\sqrt{s} = 1.96$
  TeV}'',} \textit{ Phys. Lett. B} \textbf{ 641} (2006) 415,
  \href{http://dx.doi.org/10.1016/j.physletb.2006.08.079}{\doi{10.1016/j.physletb.2006.08.079}},
\href{http://www.arXiv.org/abs/hep-ex/0605064}{\texttt{arXiv:hep-ex/0605064}}.

\bibitem{D0-2}
\hrefCMSnoop {}{{D0} Collaboration, ``{Search for a scalar or vector particle
  decaying into $\PZ \gamma$ in $\text{p} \bar{\text{p}}$ collisions at
  $\sqrt{s}=1.96$ TeV}'',} \textit{ Phys. Lett. B} \textbf{ 671} (2009) 349,
  \href{http://dx.doi.org/10.1016/j.physletb.2008.12.009}{\doi{10.1016/j.physletb.2008.12.009}},
\href{http://www.arXiv.org/abs/0806.0611}{\texttt{arXiv:0806.0611}}.

\bibitem{ATLAS-Vgamma1}
\hrefCMSnoop {}{{ATLAS Collaboration}, ``{Measurements of $\PW\gamma$ and
  $\PZ\gamma$ production in pp collisions at $\sqrt{s}=7$ TeV with the ATLAS
  detector at the LHC}'',} \textit{ Phys. Rev. D} \textbf{ 87} (2013) 112003,
  \href{http://dx.doi.org/10.1103/PhysRevD.87.112003}{\doi{10.1103/PhysRevD.87.112003}},
\href{http://www.arXiv.org/abs/1302.1283}{\texttt{arXiv:1302.1283}}.

\bibitem{ATLAS-Vgamma2}
\hrefCMSnoop {}{{ATLAS Collaboration}, ``{Search for new resonances in
  $\PW\gamma$ and $\PZ\gamma$ final states in pp collisions at $\sqrt{s}=8$ TeV
  with the ATLAS detector}'',} \textit{ Phys. Lett. B} \textbf{ 738} (2014)
  428,
  \href{http://dx.doi.org/10.1016/j.physletb.2014.10.002}{\doi{10.1016/j.physletb.2014.10.002}},
\href{http://www.arXiv.org/abs/1407.8150}{\texttt{arXiv:1407.8150}}.

\bibitem{Zgamma-lep}
\hrefCMSnoop {}{{CMS Collaboration}, ``Search for high-mass {Z$\gamma$}
  resonances in $\mathrm{e}^+\mathrm{e}^-\gamma$ and $\mu^+\mu^-\gamma$ final
  states in proton-proton collisions at $\sqrt{s}=8$ and {13\TeV}'',} \textit{
  JHEP} \textbf{ 01} (2017) 076,
  \href{http://dx.doi.org/10.1007/JHEP01(2017)076}{\doi{10.1007/JHEP01(2017)076}},
\href{http://www.arXiv.org/abs/1610.02960}{\texttt{arXiv:1610.02960}}.

\bibitem{CMS-HZg}
\hrefCMSnoop {}{{CMS Collaboration}, ``{Search for a Higgs boson decaying into
  a Z and a photon in pp collisions at $\sqrt{s} = 7$ and 8 TeV}'',} \textit{
  Phys. Lett. B} \textbf{ 726} (2013) 587,
  \href{http://dx.doi.org/10.1016/j.physletb.2013.09.057}{\doi{10.1016/j.physletb.2013.09.057}},
\href{http://www.arXiv.org/abs/1307.5515}{\texttt{arXiv:1307.5515}}.

\bibitem{ATLAS-HZg}
\hrefCMSnoop {}{{ATLAS Collaboration}, ``{Search for Higgs boson decays to a
  photon and a Z boson in pp collisions at $\sqrt{s}=7$ and 8 TeV with the
  ATLAS detector}'',} \textit{ Phys. Lett. B} \textbf{ 732} (2014) 8,
  \href{http://dx.doi.org/10.1016/j.physletb.2014.03.015}{\doi{10.1016/j.physletb.2014.03.015}},
\href{http://www.arXiv.org/abs/1402.3051}{\texttt{arXiv:1402.3051}}.

\bibitem{TRK-11-001}
\hrefCMSnoop {}{{CMS Collaboration}, ``{Description and performance of track
  and primary-vertex reconstruction with the CMS tracker}'',} \textit{ JINST}
  \textbf{ 9} (2014) P10009,
  \href{http://dx.doi.org/10.1088/1748-0221/9/10/P10009}{\doi{10.1088/1748-0221/9/10/P10009}},
\href{http://www.arXiv.org/abs/1405.6569}{\texttt{arXiv:1405.6569}}.

\bibitem{CMS:EGM-13-001}
\hrefCMSnoop {}{{CMS Collaboration}, ``Performance of electron reconstruction
  and selection with the {CMS} detector in proton-proton collisions at
  {$\sqrt{s} = 8$\TeV}'',} \textit{ JINST} \textbf{ 10} (2015) P06005,
  \href{http://dx.doi.org/10.1088/1748-0221/10/06/P06005}{\doi{10.1088/1748-0221/10/06/P06005}},
\href{http://www.arXiv.org/abs/1502.02701}{\texttt{arXiv:1502.02701}}.

\bibitem{CMS-PAPER-MUO-10-004}
\hrefCMSnoop {}{{CMS Collaboration}, ``Performance of {CMS} muon reconstruction
  in pp collision events at {$\sqrt{s} = 7$\TeV}'',} \textit{ JINST} \textbf{
  7} (2012) P10002,
  \href{http://dx.doi.org/10.1088/1748-0221/7/10/P10002}{\doi{10.1088/1748-0221/7/10/P10002}},
  \href{http://www.arXiv.org/abs/1206.4071}{\texttt{arXiv:1206.4071}}.

\bibitem{Khachatryan:2016bia}
\hrefCMSnoop {}{{CMS Collaboration}, ``{The CMS trigger system}'',} \textit{
  JINST} \textbf{ 12} (2017) P01020,
  \href{http://dx.doi.org/10.1088/1748-0221/12/01/P01020}{\doi{10.1088/1748-0221/12/01/P01020}},
\href{http://www.arXiv.org/abs/1609.02366}{\texttt{arXiv:1609.02366}}.

\bibitem{Chatrchyan:2008zzk}
\hrefCMSnoop {}{{CMS Collaboration}, ``The {CMS} experiment at the {CERN}
  {LHC}'',} \textit{ JINST} \textbf{ 3} (2008) S08004,
\href{http://dx.doi.org/10.1088/1748-0221/3/08/S08004}{\doi{10.1088/1748-0221/3/08/S08004}}.

\bibitem{Sjostrand:2007gs}
\hrefCMSnoop {}{T.~Sj{\"o}strand, S.~Mrenna, and P.~Z. Skands, ``A brief
  introduction to {PYTHIA 8.1}'',} \textit{ Comput. Phys. Commun.} \textbf{
  178} (2008) 852,
  \href{http://dx.doi.org/10.1016/j.cpc.2008.01.036}{\doi{10.1016/j.cpc.2008.01.036}},
\href{http://www.arXiv.org/abs/0710.3820}{\texttt{arXiv:0710.3820}}.

\bibitem{Monash}
\hrefCMSnoop {}{P.~Skands, S.~Carrazza, and J.~Rojo, ``{Tuning PYTHIA 8.1: the
  Monash 2013 tune}'',} \textit{ Eur. Phys. J. C} \textbf{ 74} (2014) 3024,
  \href{http://dx.doi.org/10.1140/epjc/s10052-014-3024-y}{\doi{10.1140/epjc/s10052-014-3024-y}},
\href{http://www.arXiv.org/abs/1404.5630}{\texttt{arXiv:1404.5630}}.

\bibitem{CUETP8M1}
\hrefCMSnoop {}{{CMS Collaboration}, ``{Event generator tunes obtained from
  underlying event and multiparton scattering measurements}'',} \textit{ Eur.
  Phys. J. C} \textbf{ 76} (2016) 155,
  \href{http://dx.doi.org/10.1140/epjc/s10052-016-3988-x}{\doi{10.1140/epjc/s10052-016-3988-x}},
\href{http://www.arXiv.org/abs/1512.00815}{\texttt{arXiv:1512.00815}}.

\bibitem{MADGRAPH5}
J.~Alwall\hrefCMSnoop {}{ {et~al.}, ``{MadGraph 5}: going beyond'',} \textit{
  JHEP} \textbf{ 06} (2011) 128,
  \href{http://dx.doi.org/10.1007/JHEP06(2011)128}{\doi{10.1007/JHEP06(2011)128}},
\href{http://www.arXiv.org/abs/1106.0522}{\texttt{arXiv:1106.0522}}.

\bibitem{aMCNLO}
J.~Alwall\hrefCMSnoop {}{ {et~al.}, ``{The automated computation of tree-level
  and next-to-leading order differential cross sections, and their matching to
  parton shower simulations}'',} \textit{ JHEP} \textbf{ 07} (2014) 079,
  \href{http://dx.doi.org/10.1007/JHEP07(2014)079}{\doi{10.1007/JHEP07(2014)079}},
\href{http://www.arXiv.org/abs/1405.0301}{\texttt{arXiv:1405.0301}}.

\bibitem{Ball:2014uwa}
\hrefCMSnoop {}{{NNPDF} Collaboration, ``{Parton distributions for the LHC Run
  II}'',} \textit{ JHEP} \textbf{ 04} (2015) 040,
  \href{http://dx.doi.org/10.1007/JHEP04(2015)040}{\doi{10.1007/JHEP04(2015)040}},
\href{http://www.arXiv.org/abs/1410.8849}{\texttt{arXiv:1410.8849}}.

\bibitem{Geant}
\hrefCMSnoop {}{{GEANT4} Collaboration, ``{GEANT4} --- a simulation toolkit'',}
  \textit{ Nucl. Instrum. Meth. A} \textbf{ 506} (2003) 250,
  \href{http://dx.doi.org/10.1016/S0168-9002(03)01368-8}{\doi{10.1016/S0168-9002(03)01368-8}}.

\bibitem{PFlow}
\hrefCMSnoop {}{{CMS Collaboration}, ``{Particle-flow reconstruction and global
  event description with the CMS detector}'',} \textit{ JINST} \textbf{ 12}
  (2017) P10003,
  \href{http://dx.doi.org/10.1088/1748-0221/12/10/P10003}{\doi{10.1088/1748-0221/12/10/P10003}},
\href{http://www.arXiv.org/abs/1706.04965}{\texttt{arXiv:1706.04965}}.

\bibitem{Cacciari:2008gp}
\hrefCMSnoop {}{M.~Cacciari, G.~P. Salam, and G.~Soyez, ``{The
  anti-$k_\text{t}$ jet clustering algorithm}'',} \textit{ JHEP} \textbf{ 04}
  (2008) 063,
  \href{http://dx.doi.org/10.1088/1126-6708/2008/04/063}{\doi{10.1088/1126-6708/2008/04/063}},
\href{http://www.arXiv.org/abs/0802.1189}{\texttt{arXiv:0802.1189}}.

\bibitem{Cacciari:2011ma}
\hrefCMSnoop {}{M.~Cacciari, G.~P. Salam, and G.~Soyez, ``{FastJet} user
  manual'',} \textit{ Eur. Phys. J. C} \textbf{ 72} (2012) 1896,
  \href{http://dx.doi.org/10.1140/epjc/s10052-012-1896-2}{\doi{10.1140/epjc/s10052-012-1896-2}},
\href{http://www.arXiv.org/abs/1111.6097}{\texttt{arXiv:1111.6097}}.

\bibitem{Sirunyan:2018fpa}
\hrefCMSnoop {}{{CMS Collaboration}, ``Performance of the {CMS} muon detector
  and muon reconstruction with proton-proton collisions at $\sqrt{s}=$ 13
  {TeV}'',} \textit{ JINST} \textbf{ 13} (2018) P06015,
  \href{http://dx.doi.org/10.1088/1748-0221/13/06/P06015}{\doi{10.1088/1748-0221/13/06/P06015}},
\href{http://www.arXiv.org/abs/1804.04528}{\texttt{arXiv:1804.04528}}.

\bibitem{Cacciari:2007fd}
\hrefCMSnoop {}{M.~Cacciari and G.~P. Salam, ``{Pileup subtraction using jet
  areas}'',} \textit{ Phys. Lett. B} \textbf{ 659} (2008) 119,
  \href{http://dx.doi.org/10.1016/j.physletb.2007.09.077}{\doi{10.1016/j.physletb.2007.09.077}},
\href{http://www.arXiv.org/abs/0707.1378}{\texttt{arXiv:0707.1378}}.

\bibitem{BDT1}
\hrefCMSnoop {}{H.-J. Yang, B.~P. Roe, and J.~Zhu, ``{Studies of boosted
  decision trees for MiniBooNE particle identification}'',} \textit{ Nucl.
  Instrum. Meth. A} \textbf{ 555} (2005) 370,
  \href{http://dx.doi.org/10.1016/j.nima.2005.09.022}{\doi{10.1016/j.nima.2005.09.022}},
\href{http://www.arXiv.org/abs/physics/0508045}{\texttt{arXiv:physics/0508045}}.

\bibitem{BDT2}
\href {http://pos.sissa.it/archive/conferences/050/040/ACAT_040.pdf}{H.~Voss,
  A.~H{\"o}cker, J.~Stelzer, and F.~Tegenfeldt, ``{TMVA}, the toolkit for
  multivariate data analysis with {ROOT}'',} in \textit{ XIth International
  Workshop on Advanced Computing and Analysis Techniques in Physics Research
  (ACAT)}, p.~40.
\newblock 2007.
\newblock
  \href{http://www.arXiv.org/abs/physics/0703039}{\texttt{arXiv:physics/0703039}}.
\newblock
[PoS(ACAT)040].

\bibitem{CMS-EGM-14-001}
\hrefCMSnoop {}{{CMS Collaboration}, ``Performance of photon reconstruction and
  identification with the {CMS} detector in proton-proton collisions at
  {$\sqrt{s} = 8$ TeV}'',} \textit{ JINST} \textbf{ 10} (2015) P08010,
  \href{http://dx.doi.org/10.1088/1748-0221/10/08/P08010}{\doi{10.1088/1748-0221/10/08/P08010}},
\href{http://www.arXiv.org/abs/1502.02702}{\texttt{arXiv:1502.02702}}.

\bibitem{Chatrchyan:2011ds}
\hrefCMSnoop {}{{CMS Collaboration}, ``{Determination of jet energy calibration
  and transverse momentum resolution in CMS}'',} \textit{ JINST} \textbf{ 6}
  (2011) P11002,
  \href{http://dx.doi.org/10.1088/1748-0221/6/11/P11002}{\doi{10.1088/1748-0221/6/11/P11002}},
\href{http://www.arXiv.org/abs/1107.4277}{\texttt{arXiv:1107.4277}}.

\bibitem{Khachatryan:2016kdb}
\hrefCMSnoop {}{{CMS Collaboration}, ``{Jet energy scale and resolution in the
  CMS experiment in pp collisions at 8 TeV}'',} \textit{ JINST} \textbf{ 12}
  (2017) P02014,
  \href{http://dx.doi.org/10.1088/1748-0221/12/02/P02014}{\doi{10.1088/1748-0221/12/02/P02014}},
\href{http://www.arXiv.org/abs/1607.03663}{\texttt{arXiv:1607.03663}}.

\bibitem{Ellis:2009su}
\hrefCMSnoop {}{S.~D. Ellis, C.~K. Vermilion, and J.~R. Walsh, ``Techniques for
  improved heavy particle searches with jet substructure'',} \textit{ Phys.
  Rev. D} \textbf{ 80} (2009) 051501,
  \href{http://dx.doi.org/10.1103/PhysRevD.80.051501}{\doi{10.1103/PhysRevD.80.051501}},
\href{http://www.arXiv.org/abs/0903.5081}{\texttt{arXiv:0903.5081}}.

\bibitem{Ellis:2009me}
\hrefCMSnoop {}{S.~D. Ellis, C.~K. Vermilion, and J.~R. Walsh, ``Recombination
  algorithms and jet substructure: pruning as a tool for heavy particle
  searches'',} \textit{ Phys. Rev. D} \textbf{ 81} (2010) 094023,
  \href{http://dx.doi.org/10.1103/PhysRevD.81.094023}{\doi{10.1103/PhysRevD.81.094023}},
\href{http://www.arXiv.org/abs/0912.0033}{\texttt{arXiv:0912.0033}}.

\bibitem{CA}
\hrefCMSnoop {}{Y.~L. Dokshitzer, G.~D. Leder, S.~Moretti, and B.~R. Webber,
  ``{Better jet clustering algorithms}'',} \textit{ JHEP} \textbf{ 08} (1997)
  001,
  \href{http://dx.doi.org/10.1088/1126-6708/1997/08/001}{\doi{10.1088/1126-6708/1997/08/001}},
\href{http://www.arXiv.org/abs/hep-ph/9707323}{\texttt{arXiv:hep-ph/9707323}}.

\bibitem{Khachatryan:2014vla}
\hrefCMSnoop {}{{CMS Collaboration}, ``{Identification techniques for highly
  boosted W bosons that decay into hadrons}'',} \textit{ JHEP} \textbf{ 12}
  (2014) 017,
  \href{http://dx.doi.org/10.1007/JHEP12(2014)017}{\doi{10.1007/JHEP12(2014)017}},
\href{http://www.arXiv.org/abs/1410.4227}{\texttt{arXiv:1410.4227}}.

\bibitem{Nsubjettiness}
\hrefCMSnoop {}{J.~Thaler and K.~Van~Tilburg, ``Identifying boosted objects
  with {$N$}-subjettiness'',} \textit{ JHEP} \textbf{ 03} (2011) 015,
  \href{http://dx.doi.org/10.1007/JHEP03(2011)015}{\doi{10.1007/JHEP03(2011)015}},
\href{http://www.arXiv.org/abs/1011.2268}{\texttt{arXiv:1011.2268}}.

\bibitem{subjet-axes}
\hrefCMSnoop {}{J.~Thaler and K.~Van~Tilburg, ``Maximizing boosted top
  identification by minimizing {$N$}-subjettiness'',} \textit{ JHEP} \textbf{
  02} (2012) 093,
  \href{http://dx.doi.org/10.1007/JHEP02(2012)093}{\doi{10.1007/JHEP02(2012)093}},
\href{http://www.arXiv.org/abs/1108.2701}{\texttt{arXiv:1108.2701}}.

\bibitem{JME-16-003}
\href {http://cds.cern.ch/record/2256875}{{CMS Collaboration}, ``{Jet
  algorithms performance in 13 TeV data}'',} CMS Physics Analysis Summary
  CMS-PAS-JME-16-003, 2017.

\bibitem{BTV}
\hrefCMSnoop {}{{CMS Collaboration}, ``{Identification of b-quark jets with the
  CMS experiment}'',} \textit{ JINST} \textbf{ 8} (2013) P04013,
  \href{http://dx.doi.org/10.1088/1748-0221/8/04/P04013}{\doi{10.1088/1748-0221/8/04/P04013}},
\href{http://www.arXiv.org/abs/1211.4462}{\texttt{arXiv:1211.4462}}.

\bibitem{Sirunyan:2017ezt}
\hrefCMSnoop {}{{CMS Collaboration}, ``Identification of heavy-flavour jets
  with the {CMS} detector in pp collisions at 13 {TeV}'',} \textit{ JINST}
  \textbf{ 13} (2018) P05011,
  \href{http://dx.doi.org/10.1088/1748-0221/13/05/P05011}{\doi{10.1088/1748-0221/13/05/P05011}},
\href{http://www.arXiv.org/abs/1712.07158}{\texttt{arXiv:1712.07158}}.

\bibitem{Aad:2015mzg}
\hrefCMSnoop {}{{ATLAS Collaboration}, ``{Search for strong gravity in multijet
  final states produced in pp collisions at $\sqrt{s} =$ 13 TeV using the ATLAS
  detector at the LHC}'',} \textit{ JHEP} \textbf{ 03} (2016) 026,
  \href{http://dx.doi.org/10.1007/JHEP03(2016)026}{\doi{10.1007/JHEP03(2016)026}},
\href{http://www.arXiv.org/abs/1512.02586}{\texttt{arXiv:1512.02586}}.

\bibitem{Aaltonen:2008dn}
\hrefCMSnoop {}{{CDF} Collaboration, ``{Search for new particles decaying into
  dijets in proton-antiproton collisions at $\sqrt{s} = 1.96$ TeV}'',} \textit{
  Phys. Rev. D} \textbf{ 79} (2009) 112002,
  \href{http://dx.doi.org/10.1103/PhysRevD.79.112002}{\doi{10.1103/PhysRevD.79.112002}},
\href{http://www.arXiv.org/abs/0812.4036}{\texttt{arXiv:0812.4036}}.

\bibitem{UA2-1}
\hrefCMSnoop {}{{UA2} Collaboration, ``A measurement of two-jet decays of the
  {$W$} and {$Z$} bosons at the {CERN} ${\bar{p} p}$ collider'',} \textit{ Z.
  Phys. C} \textbf{ 49} (1991) 17,
\href{http://dx.doi.org/10.1007/BF01570793}{\doi{10.1007/BF01570793}}.

\bibitem{F-test}
\hrefCMSnoop {}{R.~A. Fisher, ``{On the interpretation of $\chi^2$ from
  contingency tables, and the calculation of $p$}'',} \textit{ J. Roy. Stat.
  Soc.} \textbf{ 85} (1922) 87,
  \href{http://dx.doi.org/10.2307/2340521}{\doi{10.2307/2340521}}.

\bibitem{Khachatryan:2016hje}
\hrefCMSnoop {}{{CMS Collaboration}, ``{Search for Resonant Production of
  High-Mass Photon Pairs in Proton-Proton Collisions at $\sqrt s$ =8 and 13
  TeV}'',} \textit{ Phys. Rev. Lett.} \textbf{ 117} (2016) 051802,
  \href{http://dx.doi.org/10.1103/PhysRevLett.117.051802}{\doi{10.1103/PhysRevLett.117.051802}},
\href{http://www.arXiv.org/abs/1606.04093}{\texttt{arXiv:1606.04093}}.

\bibitem{Garwood}
\hrefCMSnoop {}{F.~Garwood, ``Fiducial limits for the {Poisson}
  distribution'',} \textit{ Biometrika} \textbf{ 28} (1936) 437,
  \href{http://dx.doi.org/10.1093/biomet/28.3-4.437}{\doi{10.1093/biomet/28.3-4.437}}.

\bibitem{CrystalBallRef}
\href {http://www.slac.stanford.edu/pubs/slacreports/ slac-r-236.html}{M.~J.
  Oreglia, ``{A study of the reactions $\psi^\prime \to \gamma \gamma
  \psi$}''}.
\newblock PhD thesis, Stanford University, 1980.
\newblock {{SLAC} Report {SLAC-R-236}}.

\bibitem{morphing}
\hrefCMSnoop {}{A.~L. Read, ``{Linear interpolation of histograms}'',} \textit{
  Nucl. Instrum. Meth. A} \textbf{ 425} (1999) 357,
\href{http://dx.doi.org/10.1016/S0168-9002(98)01347-3}{\doi{10.1016/S0168-9002(98)01347-3}}.

\bibitem{LUM-17-001}
\href {http://cds.cern.ch/record/2257069}{{CMS Collaboration}, ``{CMS}
  luminosity measurements for the 2016 data-taking period'',} CMS Physics
  Analysis Summary CMS-PAS-LUM-17-001, 2017.

\bibitem{PDF4LHC}
\hrefCMSnoop {}{J.~Butterworth {et~al.}, ``{PDF4LHC recommendations for LHC Run
  II}'',} \textit{ J. Phys. G} \textbf{ 43} (2016) 023001,
  \href{http://dx.doi.org/10.1088/0954-3899/43/2/023001}{\doi{10.1088/0954-3899/43/2/023001}},
\href{http://www.arXiv.org/abs/1510.03865}{\texttt{arXiv:1510.03865}}.

\bibitem{inelastic}
\hrefCMSnoop {}{{ATLAS Collaboration}, ``Measurement of the inelastic
  proton-proton cross section at {$\sqrt{s} = 13$ TeV} with the {ATLAS}
  detector at the {LHC}'',} \textit{ Phys. Rev. Lett.} \textbf{ 117} (2016)
  182002,
  \href{http://dx.doi.org/10.1103/PhysRevLett.117.182002}{\doi{10.1103/PhysRevLett.117.182002}},
\href{http://www.arXiv.org/abs/1606.02625}{\texttt{arXiv:1606.02625}}.

\bibitem{Gross}
\hrefCMSnoop {}{G.~Cowan, K.~Cranmer, E.~Gross, and O.~Vitells, ``{Asymptotic
  formulae for likelihood-based tests of new physics}'',} \textit{ Eur. Phys.
  J. C} \textbf{ 71} (2011) 1554,
  \href{http://dx.doi.org/10.1140/epjc/s10052-011-1554-0}{\doi{10.1140/epjc/s10052-011-1554-0}},
  \href{http://www.arXiv.org/abs/1007.1727}{\texttt{arXiv:1007.1727}}.
[Erratum: \DOI{10.1140/epjc/s10052-013-2501-z}].

\bibitem{Junk:1999kv}
\hrefCMSnoop {}{T.~Junk, ``{Confidence level computation for combining searches
  with small statistics}'',} \textit{ Nucl. Instrum. Meth. A} \textbf{ 434}
  (1999) 435,
  \href{http://dx.doi.org/10.1016/S0168-9002(99)00498-2}{\doi{10.1016/S0168-9002(99)00498-2}},
\href{http://www.arXiv.org/abs/hep-ex/9902006}{\texttt{arXiv:hep-ex/9902006}}.

\bibitem{Read:2002hq}
\hrefCMSnoop {}{A.~L. Read, ``Presentation of search results: the {$CL_s$}
  technique'',} in \textit{ Durham IPPP Workshop: Advanced Statistical
  Techniques in Particle Physics}, p.~2693.
\newblock Durham, UK, March, 2002.
\newblock [J. Phys. G 28 (2002) 2693].
  \href{http://dx.doi.org/10.1088/0954-3899/28/10/313}{\doi{10.1088/0954-3899/28/10/313}}.

\bibitem{ATL-PHYS-PUB-2011-011}
\href {https://cdsweb.cern.ch/record/1379837}{{ATLAS and CMS Collaborations},
  ``{Procedure for the LHC Higgs boson search combination in Summer 2011}'',}
  ATL-PHYS-PUB-2011-011, {CMS NOTE-2011/005}, 2011.

\bibitem{PDG}
\hrefCMSnoop {}{{Particle Data Group} Collaboration, ``{Review of Particle
  Physics}'',} \textit{ Chin. Phys. C} \textbf{ 40} (2016) 100001,
\href{http://dx.doi.org/10.1088/1674-1137/40/10/100001}{\doi{10.1088/1674-1137/40/10/100001}}.

\end{thebibliography}\endgroup
\cleardoublepage \appendix\section{The CMS Collaboration \label{app:collab}}\begin{sloppypar}\hyphenpenalty=5000\widowpenalty=500\clubpenalty=5000\vskip\cmsinstskip
\textbf{Yerevan~Physics~Institute,~Yerevan,~Armenia}\\*[0pt]
A.M.~Sirunyan, A.~Tumasyan
\vskip\cmsinstskip
\textbf{Institut~f\"{u}r~Hochenergiephysik,~Wien,~Austria}\\*[0pt]
W.~Adam, F.~Ambrogi, E.~Asilar, T.~Bergauer, J.~Brandstetter, E.~Brondolin, M.~Dragicevic, J.~Er\"{o}, A.~Escalante~Del~Valle, M.~Flechl, M.~Friedl, R.~Fr\"{u}hwirth\cmsAuthorMark{1}, V.M.~Ghete, J.~Grossmann, J.~Hrubec, M.~Jeitler\cmsAuthorMark{1}, A.~K\"{o}nig, N.~Krammer, I.~Kr\"{a}tschmer, D.~Liko, T.~Madlener, I.~Mikulec, E.~Pree, N.~Rad, H.~Rohringer, J.~Schieck\cmsAuthorMark{1}, R.~Sch\"{o}fbeck, M.~Spanring, D.~Spitzbart, W.~Waltenberger, J.~Wittmann, C.-E.~Wulz\cmsAuthorMark{1}, M.~Zarucki
% ...
\vskip\cmsinstskip
% ...
1:~Also at~Vienna~University~of~Technology,~Vienna,~Austria\\
% ...
\end{sloppypar}
\end{document}